\documentclass[twocolumn,showpacs,amsmath,amssymb,superscriptaddress]{revtex4-1}
\usepackage{dcolumn}
\usepackage{color}
\usepackage{graphicx}
\usepackage{amsmath}
\usepackage{multirow,bm,url,tabularx,longtable,hyperref}

\newcommand\+{\dagger}

\usepackage{longtable}	
\begin{document}

\title{Structure of odd-odd Cs isotopes within the interacting boson-fermion-fermion model based on the Gogny-D1M energy density functional}

\author{K.~Nomura}
\email{knomura@phy.hr}
\affiliation{Department of Physics, Faculty of Science, University of
Zagreb, HR-10000 Zagreb, Croatia}

\author{R.~Rodr\'iguez-Guzm\'an}
\affiliation{Physics Department, Kuwait University, 13060 Kuwait, Kuwait}

\author{L.~M.~Robledo}
\affiliation{Departamento de F\'\i sica Te\'orica, Universidad
Aut\'onoma de Madrid, E-28049 Madrid, Spain}

\affiliation{Center for Computational Simulation,
Universidad Polit\'ecnica de Madrid,
Campus de Montegancedo, Boadilla del Monte, 28660-Madrid, Spain
}

\date{\today}

\begin{abstract}
The spectroscopic properties of the odd-odd isotopes $^{124-132}$Cs 
have been studied within the interacting boson-fermion-fermion model 
based on the Gogny-D1M  energy density functional framework. Major 
ingredients to build the interacting boson-fermion-fermion Hamiltonian, 
such as the ($\beta,\gamma$)-deformation energy surfaces for the 
even-even core nuclei $^{124-132}$Xe as well as single-particle 
energies and occupation probabilities of the odd nucleons, have been 
computed  microscopically with the constrained Hartree-Fock-Bogoliubov 
method. A few coupling constants of the boson-fermion and residual 
neutron-proton interactions are fitted to reproduce with a reasonable 
accuracy the experimental excitation energy of the low-lying levels of 
the odd-mass and odd-odd nuclei. The method is applied to describe the 
low-energy low-spin spectra of the odd-odd Cs nuclei and the band 
structures of higher-spin higher-energy states, mainly based  on the 
$(\nu h_{11/2})^{-1}\otimes\pi h_{11/2}$ configuration. Many of those 
odd-odd Cs nuclei have been identified as candidates for exhibiting 
chiral doublet bands. 
\end{abstract}


\maketitle


\section{Introduction}


A better understanding of the spectroscopic properties of atomic nuclei 
with an odd number of nucleons still remains a major challenge for both 
experimental and theoretical low-energy nuclear physics. The existence 
of an unpaired nucleon in the nucleus implies the observation of many 
new effects in nuclear dynamics like the weakening of pairing 
correlations, the increase of level densities around the Fermi level, 
polarization of collective degrees of freedom, breaking of time 
reversal symmetry in the intrinsic wave function, and a long list of 
etc. As a consequence, the microscopic description of an odd-A system 
is far more challenging than in the traditional even-even case 
\cite{RS,bender2003,Rob19}. This is manifest in the much slower 
progress in the implementation of symmetry restoration in odd-A nuclei 
\cite{bally2014,borrajo2016}. In addition, the quantitative side is 
strongly affected by tiny details of the nuclear interaction, making 
this kind of systems the perfect test ground to analyze the suitability 
of new or existing proposals for effective nuclear 
interactions/functionals, see \cite{Dobaczewski2015} for a recent 
analysis focusing  on superheavy nuclei. 
Detailed spectroscopic studies 
of odd-mass and/or odd-odd nuclei, have already been carried out  using 
microscopic approaches such as the large-scale shell model 
\cite{caurier2005} and the symmetry-projected generator coordinate 
method (GCM) \cite{bally2014,borrajo2016}. See \cite{RS} for a general 
introduction to the latter method. From a computational point of view, 
systematic applications of these approaches are very demanding, if not 
impossible, for heavy nuclei, especially when a large number of valence 
nucleons are involved and/or multiple shape degrees of freedom have to 
be taken into account in the generator coordinate method (GCM) ansatz. 

To overcome these difficulties we proposed in 
Ref.~\cite{nomura2019dodd} (in a study of odd-A Au and Pt and odd-odd Au 
isotopes) to perform constrained Hartree-Fock-Bogoliubov (HFB) 
calculations based on the Gogny \cite{Gogny} energy density functional (EDF) with the parametrization D1M \cite{D1M}, to obtain energy 
surfaces as functions of the ($\beta,\gamma$) quadrupole deformation 
parameters for the neighboring even-even Pt nuclei. The single-particle 
energies and occupation numbers were computed for the odd neutron and 
odd proton in the odd-mass Au and Pt as well as odd-odd Au isotopes. 
Those quantities were then used, as a microscopic input, to completely 
determine the interacting boson model (IBM) \cite{IBM} Hamiltonian 
for the even-even nucleus and most of the parameters of the 
different boson-fermion coupling terms present in the interacting boson-fermion model (IBFM) \cite{iachello1979,scholten1985,IBFM} and 
the interacting boson-fermion-fermion model (IBFFM) \cite{brant1984,IBFM} Hamiltonians for the odd-A and odd-odd systems, respectively. Only 
a few coupling constants of the boson-fermion and the residual 
neutron-proton interaction terms were treated as free parameters. These 
parameters were determined so as to reproduce reasonably well the 
experimental low-lying energy levels of the odd-mass and odd-odd 
nuclei. Though the method involves a few phenomenological parameters, 
it allows to study simultaneously the spectroscopy of even-even, 
odd-mass, and odd-odd nuclei within a unified framework. The method 
reduces significantly the computational cost associated with those 
calculations and  provides the possibility of studying heavy odd and 
odd-odd nuclei irrespective of their location at the chart of nuclides.

In this work, we consider the spectroscopic properties of the odd-odd 
nuclei $^{124-132}$Cs, using the theoretical framework developed in 
Ref.~\cite{nomura2019dodd}.  
The reason for the choice of nuclei is that the  $A\approx$130  mass 
region exhibits a wide variety of structural phenomena. A variety of 
theoretical models suggested the existence of triaxially-deformed 
and/or $\gamma$-soft shapes for even-even systems in this mass region 
\cite{CASTEN1985,sevrin1987,yan1993,vogel1996,mizusaki1997,yoshinaga2004,li2010,nomura2012tri}. 
A gradual transition, from $\gamma$-soft to nearly spherical shapes, 
has also been identified \cite{cejnar2010} while several nuclei, such 
as $^{134}$Ba \cite{casten2000} and $^{128}$Xe \cite{coquard2009}, are 
suggested to display features of the E(5) critical-point symmetry 
\cite{iachello2000} of the phase transition. In some odd-odd Cs 
isotopes, most notably in $^{128}$Cs, chiral doublet bands 
\cite{frauendorf1997} have been observed 
\cite{koike2004,grodner2006,starosta2017,grodner2018}. Those bands are 
associated with nearly degenerate energy levels with equal spins and 
characteristic electromagnetic properties. The high-spin level structure in  
the odd-odd nuclei in the mass $A\approx 130$ region, in particular the role of the 
$(\nu h_{11/2})^{-1}\otimes\pi h_{11/2}$ (neutron hole coupled with proton) 
or $\nu h_{11/2}\otimes\pi h_{11/2}$ configuration in forming the chiral bands, has been studied by 
various theoretical approaches, and we particularly mention the IBFFM \cite{brant2004} and 
shell model \cite{yoshinaga2006,higashiyama2005,higashiyama2007,higashiyama2013} 
calculations. 
Furthermore, this mass 
region represents a challenging testing ground to examine the 
predictive power of nuclear models for fundamental  processes, such as 
$\beta$-decay and double-$\beta$ decay 
\cite{zuffi2003,brant2006,mardones2016,engel2017}. Previous 
phenomenological IBFM and IBFFM spectroscopic studies \cite{arias1985} 
were also carried out for nuclei in the same mass region considered in 
this work.

The paper is organized as follows. In Sec.~\ref{sec:model}, we outline 
the theoretical framework used in this study. We begin 
Sec.~\ref{sec:results}, with a brief discussion of the results obtained 
for the even-even core nucleus $^{124}$Xe as well as the odd-N and 
odd-Z nuclei $^{123}$Xe and $^{125}$Cs. In the same section, we discuss 
the low-energy spectra obtained for the odd-odd systems $^{124-132}$Cs. 
Moreover, we pay attention to the band structures of higher-spin states  
to identify features of chirality in some of the considered odd-odd Cs 
isotopes. Finally, Sec.~\ref{sec:summary} is devoted to the concluding 
remarks.


\section{Theoretical framework\label{sec:model}}



\subsection{IBFFM-2 Hamiltonian}


Within the employed theoretical scheme, the 
low-lying structure of the even-even-core nucleus is described in terms 
of the IBM \cite{IBM}, where correlated pairs 
of valence nucleons are represented by bosonic degrees of freedom 
\cite{OAI}. In the IBFM, one unpaired nucleon is explicitly included as an additional degree of 
freedom to the boson space \cite{iachello1979,scholten1985,IBFM} to handle odd-mass systems. 
The IBFFM represents a 
further extension of the IBFM to odd-odd systems that includes, one 
unpaired neutron and one unpaired proton \cite{brant1984,IBFM}. 
As in our previous study for odd-odd Au isotopes \cite{nomura2019dodd}, 
we have used a version of the IBFFM that distinguishes between neutron and 
proton degrees of freedom (denoted hereafter as IBFFM-2). 
The IBFFM-2 Hamiltonian reads 
\begin{equation}
\label{eq:ham}
 \hat H_\text{} = \hat H_\text{B} + \hat H_\text{F}^\nu + \hat
 H_\text{F}^\pi + \hat
  H_\text{BF}^\nu + H_\text{BF}^\pi + \hat V_\text{res}. 
\end{equation}
where the first term represents  the neutron-proton IBM (IBM-2)
Hamiltonian \cite{OAI}  that describes the  even-even core nuclei
($^{124,126,128,130,132}$Xe). The second (third) term is the
Hamiltonian for an odd neutron (proton). The fourth (fifth) term
corresponds to the interaction Hamiltonian  describing the coupling of the odd
neutron (proton) to the IBM-2 core. The last term 
in Eq.~(\ref{eq:ham}) is the residual interaction between
the odd neutron and the odd proton. 

For the boson-core Hamiltonian $\hat H_\text{B}$ in Eq.~(\ref{eq:ham}) 
the standard IBM-2 Hamiltonian has been  adopted: 
\begin{equation}
\label{eq:ibm2}
 \hat H_{\text{B}} = \epsilon(\hat n_{d_\nu} + \hat n_{d_\pi})+\kappa\hat
  Q_{\nu}\cdot\hat Q_{\pi}
\end{equation}
where $\hat n_{d_\rho}=d^\dagger_\rho\cdot\tilde d_{\rho}$ ($\rho=\nu,\pi$) 
is the $d$-boson number operator while $\hat Q_\rho=d_\rho^\dagger s_\rho +
s_\rho^\dagger\tilde d_\rho^\dagger + \chi_\rho(d^\dagger_\rho\times\tilde
d_\rho)^{(2)}$ is the quadrupole operator. 
The parameters of the Hamiltonian are 
$\epsilon$, $\kappa$, $\chi_\nu$, and $\chi_\pi$. 
The doubly-magic nucleus $^{100}$Sn is taken as the inert 
core for the boson space. We have followed the standard way of counting  the
number of bosons, i.e., the numbers of neutron  
$N_{\nu}$ and proton  $N_{\pi}$ bosons equal the numbers of 
neutron-hole and proton-particle pairs, respectively.
As a consequence, $N_\pi=2$ and $N_{\nu}=6$, 5, 4, 3 and 2 
for  $^{124,126,128,130,132}$Xe, respectively.

In Eq.~(\ref{eq:ham}), the Hamiltonian for the odd nucleon, i.e., $\hat 
H_{\text F}^{\rho}$ takes the form 
\begin{equation}
\label{one-body}
 \hat H_\text{F}^\rho = -\sum_{j_\rho}\epsilon_{j_\rho}\sqrt{2j_\rho+1}
  (a_{j_\rho}^\dagger\times\tilde a_{j_\rho})^{(0)}
\end{equation}
where $\epsilon_{j_\nu}$ ($\epsilon_{j_\pi}$) and $j_\nu$ ($j_\pi$) 
stand for the single-particle energy and the angular momentum of the 
unpaired neutron (proton). On the other hand, $a_{j_\rho}^{(\dagger)}$ 
($a_{j_\rho}$) represents the fermion creation (annihilation) operator 
while $\tilde a_{j_\rho}$ is defined as $\tilde 
a_{jm}=(-1)^{j-m}a_{j-m}$. For the fermion valence space, we have taken 
into account the full neutron and proton major shell $N,Z=50-82$, that 
include the $3s_{1/2}$, $2d_{3/2}$, $2d_{5/2}$, $1g_{7/2}$, and 
$1h_{11/2}$ orbitals. 

For the boson-fermion interaction term, $\hat H_{\rm BF}^\rho$ in
Eq.~(\ref{eq:ham}), we employ the form that has been formulated within a simple generalized 
seniority scheme \cite{scholten1985,IBFM}:  
\begin{equation}
\label{eq:ham-bf}
 \hat H_\text{BF}^\rho = \Gamma_\rho\hat Q_{\rho'}\cdot\hat q_{\rho} 
+
  \Lambda_\rho\hat V_{\rho'\rho} + A_\rho\hat n_{d_{\rho}}\hat n_{\rho}
\end{equation}
where $\rho'\neq\rho$, and the first, second, and third terms are the 
quadrupole dynamical, exchange, and monopole terms, respectively. The 
strength parameters of the interaction Hamiltonian are denoted by 
$\Gamma_\rho$, $\Lambda_\rho$, and $A_{\rho}$. As in previous studies 
\cite{scholten1985,arias1986}, we have assumed that both the dynamical 
and exchange terms are dominated by the interaction between unlike 
particles, i.e., between the odd neutron and proton bosons and between 
the odd proton and neutron bosons. We also assume that for the monopole 
term the interaction between like-particles, i.e., between the odd 
neutron and neutron bosons and between the odd proton and proton 
bosons, plays a dominant role. In Eq.~(\ref{eq:ham-bf}), $\hat Q_\rho$ 
is the bosonic quadrupole operator identical to the one in the IBM-2 
Hamiltonian in Eq.~(\ref{eq:ibm2}) with the same value of the parameter 
$\chi_{\rho}$. The fermionic quadrupole operator $\hat q_\rho$ reads 
\begin{equation}
\hat q_\rho=\sum_{j_\rho j'_\rho}\gamma_{j_\rho j'_\rho}(a^\+_{j_\rho}\times\tilde
a_{j'_\rho})^{(2)},
\end{equation} 
where $\gamma_{j_\rho
j'_\rho}=(u_{j_\rho}u_{j'_\rho}-v_{j_\rho}v_{j'_\rho})Q_{j_\rho
j'_\rho}$ and  $Q_{j_\rho j'_\rho}=\langle
l\frac{1}{2}j_{\rho}||Y^{(2)}||l'\frac{1}{2}j'_{\rho}\rangle$ represents
the matrix element of the fermionic 
quadrupole operator in the considered single-particle basis.
The exchange term $\hat V_{\rho'\rho}$ in Eq.~(\ref{eq:ham-bf}) reads 
\begin{eqnarray}
\label{eq:exchange}
 \hat V_{\rho'\rho} =& -(s_{\rho'}^\+\tilde d_{\rho'})^{(2)}
\cdot
\Bigg\{
\sum_{j_{\rho}j'_{\rho}j''_{\rho}}
\sqrt{\frac{10}{N_\rho(2j_{\rho}+1)}}\beta_{j_{\rho}j'_{\rho}}\beta_{j''_{\rho}j_{\rho}} \nonumber \\
&:((d_{\rho}^\+\times\tilde a_{j''_\rho})^{(j_\rho)}\times
(a_{j'_\rho}^\+\times\tilde s_\rho)^{(j'_\rho)})^{(2)}:
\Bigg\} + (H.c.), \nonumber \\
\end{eqnarray}
with $\beta_{j_{\rho}j'_{\rho}}=(u_{j_{\rho}}v_{j'_{\rho}}+v_{j_{\rho}}u_{j'_{\rho}})Q_{j_{\rho}j'_{\rho}}$. 
In the second line of the above equation the standard notation $:(\cdots):$ indicates normal ordering. 
For the monopole term, the number operator for the odd fermion is
expressed as $\hat
n_{\rho}=\sum_{j_{\rho}}(-\sqrt{2j_{\rho}+1})(a^\+_{j_\rho}\times\tilde 
a_{j_\rho})^{(0)}$.

Finally, we adopted the following form of the 
residual neutron-proton interaction $\hat V_{\text{res}}$ 
\begin{equation}
\label{eq:res}
 \hat V_{\text{res}}=4\pi u_{\rm D}\delta({\bf r_\nu}-{\bf
  r_\pi})
+u_{\rm T}\Bigg\{\frac{3(\sigma_{\nu}\cdot{\bf
r_{\nu\pi}})(\sigma_{\pi}\cdot{\bf
r_{\nu\pi}})}{r^2_{\nu\pi}}-\sigma_{\nu}\cdot\sigma_{\pi}\Bigg\}, 
\end{equation}
where the first and second terms denote the delta and tensor 
interactions, respectively. We have found that these two terms are
enough to provide a reasonable description 
of the low-lying states in the considered  odd-odd nuclei. 
Note that by definition ${\bf r_{\nu\pi}}={\bf r_{\nu}}-{\bf r_{\pi}}$ and that 
$u_{\rm D}$ and $u_{\rm T}$ are the parameters of this term. 
Furthermore, the matrix element $V_\text{res}'$ of the residual interaction 
$\hat V_\text{res}$ can be expressed as
\cite{yoshida2013}: 
\begin{align}
\label{eq:vres}
V_\text{res}'
&= (u_{j_\nu'} u_{j_\pi'} u_{j_\nu} u_{j_\nu} + v_{j_\nu'} v_{j_\pi'} v_{j_\nu} v_{j_\nu})
V^{J}_{j_\nu' j_\pi' j_\nu j_\pi}
\nonumber \\
& {} - (u_{j_\nu'}v_{j_\pi'}u_{j_\nu}v_{j_\pi} +
 v_{j_\nu'}u_{j_\pi'}v_{j_\nu}u_{j_\pi}) \nonumber \\
&\times \sum_{J'} (2J'+1)
\left\{ \begin{array}{ccc} {j_\nu'} & {j_\pi} & J' \\ {j_\nu} & {j_\pi'} & J
\end{array} \right\} 
V^{J'}_{j_\nu'j_\pi j_\nu j_\pi'}, 
\end{align}
where
\begin{equation}
V^{J}_{j_\nu'j_\pi'j_\nu j_\pi} = \langle j_\nu'j_\pi';J|\hat
 V_\text{res}|j_\nu j_\pi;J\rangle
\end{equation}
represents the matrix element between the neutron-proton pairs 
and $J$ stands
for the total angular momentum of the neutron-proton pair. 
The bracket in Eq.~(\ref{eq:vres}) represents the corresponding Racah coefficient.
The terms resulting from contractions are neglected
in Eq.~(\ref{eq:vres}), as in Ref.~\cite{morrison1981}.


\subsection{Procedure to build the IBFFM-2 Hamiltonian}


The basic ingredients of the
IBFFM-2 Hamiltonian $\hat H$ in Eq.~(\ref{eq:ham}) 
are determined as follows \cite{nomura2019dodd}: 
\begin{enumerate}
\item Once the form of the IBM-2 Hamiltonian is fixed, the parameters  
 $\epsilon$, $\kappa$, $\chi_\nu$, and $\chi_\pi$ are uniquely  
       determined \cite{nomura2008,nomura2010} by mapping the $(\beta,\gamma)$-deformation
       energy surface obtained from the constrained Gogny-D1M \cite{D1M}
       HFB calculation 
       onto the expectation value of the IBM-2 Hamiltonian in the boson
       coherent state \cite{ginocchio1980}.

\item The single-neutron Hamiltonian $\hat H_{\rm F}^{\nu}$ and the
      boson-fermion Hamiltonian $\hat H^{\nu}_{\rm BF}$ for 
      odd-N Xe isotopes are 
      built by using the procedure of
      \cite{nomura2016odd} (see also \cite{nomura2017odd-2} for further details ). 
      In those references, the  single-particle energies and 
      occupation probabilities of the odd nucleon, entering both $\hat H_{\rm F}^{\nu}$ and $\hat H^{\nu}_{\rm BF}$, 
      are obtained from  
      Gogny-D1M HFB calculations at zero deformation. The optimal values of the 
      boson-fermion interaction strengths 
       $\Gamma_\nu$, $\Lambda_\nu$, and $A_\nu$ in Eq.~(\ref{eq:ham-bf}), are 
       chosen, separately for positive and negative parity, so as to reproduce          
       with a reasonable accuracy the  experimental low-energy levels of each 
       odd-N Xe nucleus. A similar procedure has been employed to determine  the parameters 		
       $\Gamma_\pi$, $\Lambda_\pi$, and $A_\pi$ for the odd-Z Cs isotopes.

\item We use for the IBFFM-2 Hamiltonian in the odd-odd Cs the same 
      strength parameters $\Gamma_\nu$, $\Lambda_\nu$, and $A_\nu$
      ($\Gamma_\pi$, $\Lambda_\pi$, and $A_\pi$) obtained for
      the odd-N Xe (odd-Z Cs) nuclei in
      the previous step. The single-particle energies and occupation probabilities are, 
      however, computed independently for each of the studied odd-odd systems. 

\item Finally, the parameters $u_{\rm D}$ and $u_{\rm T}$, in the residual 
      interaction $\hat V_{\text{res}}$, are
      determined so as to reproduce with reasonable accuracy the low-lying
      spectra in the odd-odd nuclei under consideration. 
      For simplicity, we have taken the fixed values $u_{\rm D}=0.7$ MeV
      and $u_{\rm T}=0.02$ MeV for all the considered nuclei and 
      for both parities. 
\end{enumerate}

The  values of the IBM-2 parameters adopted for the even-even Xe 
isotopes are shown in Table~\ref{tab:ibm2para}. In particular, the sum 
$\chi_{\nu}+\chi_{\pi}$ is somewhat close to zero in many of the 
considered Xe isotopes. This indicates that these nuclei are close to 
the O(6) limit of the IBM, which is associated with $\gamma$-soft 
deformation. 

The fitted strength parameters of the boson-fermion interactions, $\hat 
H_{\rm BF}^{\rho}$, are shown in Table~\ref{tab:para-dodd}. The values 
of some of these strength parameters, i.e., $\Gamma_{\rho}$ and 
$\Lambda_{\rho}$, for a given configuration ($sdg$ or $h_{11/2}$) 
gradually change with neutron number. For the positive-parity states in 
$^{128,130,132}$Cs, the values of $\Gamma_{\pi}$ for the proton 
$h_{11/2}$ configuration ( which are fitted to the odd-mass nuclei 
$^{129,131,133}$Cs, respectively) have been modified so that the 
higher-spin positive-parity states, which are mainly composed  of the 
$(\nu h_{11/2})^{-1}\otimes\pi h_{11/2}$ configuration, become lower in 
energy. We consider a value of $\approx 0.5$ MeV for the excitation 
energy $E_{\mathrm x}$. The modified $\Gamma_{\pi}$ values, given in 
parentheses in Table~\ref{tab:para-dodd}, are also different from those 
employed for the negative-parity states.

Finally, the single-particle energies and occupation probabilities for 
the odd-odd Cs isotopes, obtained using  the Gogny-D1M  HFB 
approach, are given in Table~\ref{tab:vsq-dodd}. They are quite similar 
to the ones obtained in the case of the  odd-N Xe and odd-Z Cs nuclei. 

\begin{table}[htb!]
 \begin{center}
\caption{\label{tab:ibm2para} Parameters of the IBM-2
  Hamiltonian $\hat H_\text{B}$ for the even-even isotopes $^{124-132}$Xe.}
\begin{ruledtabular}
  \begin{tabular}{ccccc}
   & $\epsilon$ (MeV) & $\kappa$ (MeV) & $\chi_\nu$ & $\chi_\pi$ \\
\hline
$^{124}$Xe & 0.45 & $-0.336$ & 0.40 & $-0.50$ \\
$^{126}$Xe & 0.52 & $-0.323$ & 0.25 & $-0.50$ \\
$^{128}$Xe & 0.62 & $-0.315$ & 0.25 & $-0.55$ \\
$^{130}$Xe & 0.82 & $-0.308$ & 0.38 & $-0.50$ \\
$^{132}$Xe & 0.90 & $-0.250$ & 0.20 & $-0.55$ 
  \end{tabular}
  \end{ruledtabular}
 \end{center}
\end{table}


\begin{table}[htb!]
\caption{\label{tab:para-dodd} Parameters for the boson-fermion 
coupling Hamiltonians $\hat H^{\nu}_{\mathrm BF}$ and $\hat H^{\pi}_{\mathrm BF}$ 
(in MeV). These values have been adopted for describing the  odd-odd nuclei $^{124-132}$Cs. 
For the positive-parity states in $^{128,130,132}$Cs, the values of the parameter 
$\Gamma_{\pi}$ for the $h_{11/2}$ orbital are  different compared to those 
employed for the 
negative-parity states and are shown in parentheses.} 
 \begin{center}
 \begin{ruledtabular}
  \begin{tabular}{lccccccc}
   & & $\Gamma_\nu$ & $\Lambda_\nu$ & $A_\nu$ & $\Gamma_\pi$ & $\Lambda_\pi$ & $A_\pi$ \\
\hline
$^{124}$Cs & $sdg$ & 3.20 & 0.20 & $-0.14$ & 0.80 & 0.51 & $-0.80$ \\
                    & $h_{11/2}$ & 3.20 & 4.80 & $-0.20$ & 0.60 & 0.51 & $-2.2$ \\
                    \hline
$^{126}$Cs & $sdg$ & 3.00 & 0.40 & $-0.12$ & 0.80 & 0.40 & $-0.70$ \\
                    & $h_{11/2}$ & 3.00 & 1.85 & 0.00 & 1.00 & 0.50 & $-1.0$ \\
                    \hline
$^{128}$Cs & $sdg$ & 3.00 & 0.60 & $-0.28$ & 1.00 & 0.40 & $-0.70$ \\
                    & $h_{11/2}$ & 3.00 & 1.33 & 0.00 & 1.00 (2.60) & 0.50 & $-1.3$ \\
                    \hline
$^{130}$Cs & $sdg$ & 1.60 & 2.20 & $-0.30$ & 1.20 & 0.55 & $-0.80$ \\
                    & $h_{11/2}$ & 1.60 & 0.92 & $-0.48$ & 1.20 (2.40) & 0.55 & $-1.3$ \\
                    \hline
$^{132}$Cs & $sdg$ & 1.00 & 2.00 & $-0.30$ & 1.20 & 0.58 & $-0.50$ \\
                    & $h_{11/2}$ & 1.00 & 0.95 & $-0.34$ & 1.20 (3.00) & 0.58 & $-0.55$ \\
  \end{tabular}
  \end{ruledtabular}
 \end{center}
\end{table}


\begin{table*}[htb!]
\caption{\label{tab:vsq-dodd} Neutron and proton single-particle
 energies (in MeV) and occupation probabilities for the odd-odd Cs isotopes.}
 \begin{center}
 \begin{ruledtabular}
  \begin{tabular}{cccccccccccccc}
   & & $3s_{1/2}$ & $2d_{3/2}$ & $2d_{5/2}$ & $1g_{7/2}$ & $1h_{11/2}$ &
   & $3s_{1/2}$ & $2d_{3/2}$ & $2d_{5/2}$ & $1g_{7/2}$ & $1h_{11/2}$ \\
\hline
$^{124}$Cs & $\epsilon_{j_{\nu}}$ & 1.339 & 1.003 & 3.719 & 3.439 & 0.000 & $\epsilon_{j_{\pi}}$ & 2.555 & 2.476 & 0.122 & 0.000 & 3.674 \\ 
 & $v^2_{j_{\nu}}$ & 0.602 & 0.506 & 0.929 & 0.902 & 0.243 & $v^2_{j_{\pi}}$ & 0.034 & 0.047 & 0.303 & 0.352 & 0.023 \\ 
\hline
$^{126}$Cs & $\epsilon_{j_{\nu}}$ & 1.271 & 0.983 & 3.684 & 3.516 & 0.000 & $\epsilon_{j_{\pi}}$ & 2.680 & 2.525 & 0.207 & 0.000 & 3.674 \\ 
 & $v^2_{j_{\nu}}$ & 0.692 & 0.618 & 0.944 & 0.925 & 0.332 & $v^2_{j_{\pi}}$ & 0.032 & 0.047 & 0.290 & 0.362 & 0.024 \\ 
\hline
$^{128}$Cs & $\epsilon_{j_{\nu}}$ & 1.217 & 0.978 & 3.656 & 3.607 & 0.000 & $\epsilon_{j_{\pi}}$ & 2.809 & 2.580 & 0.298 & 0.000 & 3.668 \\ 
 & $v^2_{j_{\nu}}$ & 0.770 & 0.718 & 0.956 & 0.943 & 0.431 & $v^2_{j_{\pi}}$ & 0.030 & 0.046 & 0.276 & 0.373 & 0.024 \\ 
\hline
$^{130}$Cs & $\epsilon_{j_{\nu}}$ & 1.174 & 0.984 & 3.635 & 3.710 & 0.000 & $\epsilon_{j_{\pi}}$ & 2.942 & 2.642 & 0.392 & 0.000 & 3.655 \\ 
 & $v^2_{j_{\nu}}$ & 0.838 & 0.805 & 0.968 & 0.958 & 0.541 & $v^2_{j_{\pi}}$ & 0.028 & 0.045 & 0.261 & 0.384 & 0.025 \\ 
\hline
$^{132}$Cs & $\epsilon_{j_{\nu}}$ & 1.141 & 1.001 & 3.620 & 3.823 & 0.000 & $\epsilon_{j_{\pi}}$ & 3.081 & 2.712 & 0.5 & 0.000 & 3.637 \\ 
 & $v^2_{j_{\nu}}$ & 0.896 & 0.878 & 0.977 & 0.972 & 0.660 & $v^2_{j_{\pi}}$ & 0.026 & 0.044 & 0.246 & 0.395 & 0.025 
  \end{tabular}
  \end{ruledtabular}
 \end{center}
\end{table*}

Once the value of all the parameters  has been obtained, the IBFFM-2 
Hamiltonian is diagonalized in the $|L_\nu L_\pi(L);j_\nu 
j_\pi(J):I\rangle$ basis characterized by the angular momentum of the 
neutron (proton) bosons $L_\nu$ ($L_{\pi}$), the total angular momentum 
for the even-even boson core $L$ and the total angular momentum of the 
coupled system $I$. 


\subsection{Transition operators}


Using the wave functions obtained after the diagonalization of the 
IBFFM-2 Hamiltonian, the electric quadrupole (E2) and magnetic dipole 
(M1) properties can be computed. The corresponding $\hat T^{(E2)}$ and 
$\hat T^{(M1)}$ operators are given by \cite{nomura2019dodd} 
\begin{align}
\label{eq:e2}
\hat T^{(E2)}&= e_\nu^B\hat Q_\nu + e_\pi^B\hat Q_\pi
-\frac{1}{\sqrt{5}}\sum_{\rho=\nu,\pi}\sum_{j_{\rho}j'_{\rho}} \nonumber \\
&\times(u_{j_{\rho}}u_{j'_{\rho}}-v_{j_{\rho}}v_{j'_{\rho}})\langle
j'_{\rho}||e^F_{\rho}r^2Y^{(2)}||j_{\rho}\rangle(a_{j_{\rho}}^\dagger\times\tilde a_{j'_{\rho}})^{(2)},
\nonumber \\
\end{align}
and 
\begin{align}
\label{eq:m1}
\hat T^{(M1)}&=\sqrt{\frac{3}{4\pi}}
\Big\{
g_\nu^B\hat L^B_\nu + g_\pi^B\hat
L^B_\pi
-\frac{1}{\sqrt{3}}\sum_{\rho=\nu,\pi}\sum_{jj'} \nonumber \\
&\times (u_{j_{\rho}}u_{j'_{\rho}}+v_{j_{\rho}}v_{j'_{\rho}})\langle
j'_{\rho}||g_l^\rho{\bf l}+g_s^\rho{\bf s}||j_{\rho}\rangle(a_{j_{\rho}}^\dagger\times\tilde
a_{j'_{\rho}})^{(1)}
\Big\}. \nonumber \\
\end{align}
In Eq.~(\ref{eq:e2}), $e^B_\rho$ and $e^F_{\rho}$ are the effective 
charges for the boson and fermion systems. We have employed the fixed 
values $e^B_\nu=e^B_\pi=0.15$ $e$b, and $e^F_\nu=0.5$ $e$b and 
$e^F_\pi=1.5$ $e$b. In the case of the  M1 operator in 
Eq.~(\ref{eq:m1}), $g_\nu^B$ and $g_\pi^B$ are  $g$-factors for the 
neutron and proton bosons. We have also used the fixed values 
$g_\nu^B=0\,\mu_N$ and $g_\pi^B=1.0\,\mu_N$  \cite{yoshida1985,IBM}. 
For the neutron (proton) $g$-factors, the usual Schmidt values 
$g_l^\nu=0\,\mu_N$ and $g_s^\nu=-3.82\,\mu_N$ ($g_l^\pi=1.0\,\mu_N$ and 
$g_s^\pi=5.58\,\mu_N$) have been considered. Both the  proton and 
neutron $g_s$ values have been quenched  30 \%.


\section{Results and discussion \label{sec:results}}


In this section, we will briefly discuss some selected results obtained 
for even-even Xe and odd-mass Cs nuclei. The nuclei $^{124}$Xe 
(Sec.~\ref{sec:ee}), $^{123}$Xe and  $^{125}$Cs (Sec.~\ref{sec:oe}) 
will be taken as representative examples. As we are mainly interested 
in the structure of odd-odd nuclei, most of our discussions will be 
devoted to the spectroscopic results obtained for such odd-odd systems 
(Sec.~\ref{sec:oo}).

\subsection{Even-even nuclei\label{sec:ee}}


\begin{figure}[htb!]
\begin{center}
\includegraphics[width=\linewidth]{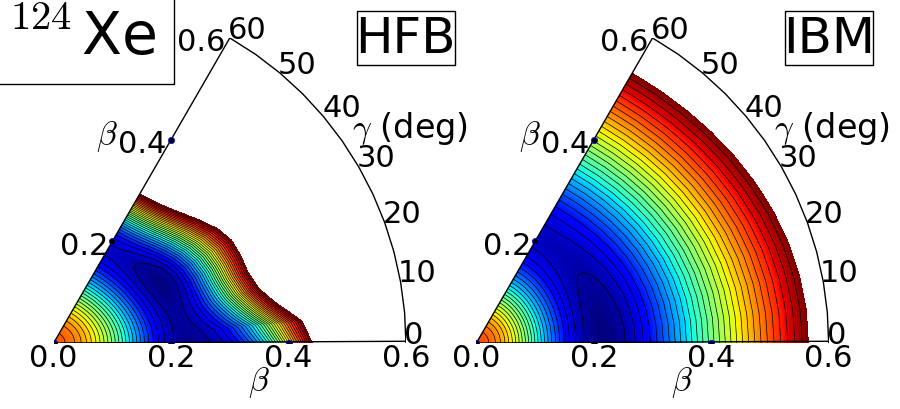}
\caption{(Color online) The Gogny-D1M and IBM-2
 $(\beta,\gamma)$-deformation energy surfaces obtained for 
 $^{124}$Xe are plotted up to 3 MeV from the global minimum. 
 The energy difference between neighboring contours is
 100 keV. } 
\label{fig:124xe-pes}
\end{center}
\end{figure}


\begin{figure}[htb!]
\begin{center}
\includegraphics[width=0.8\linewidth]{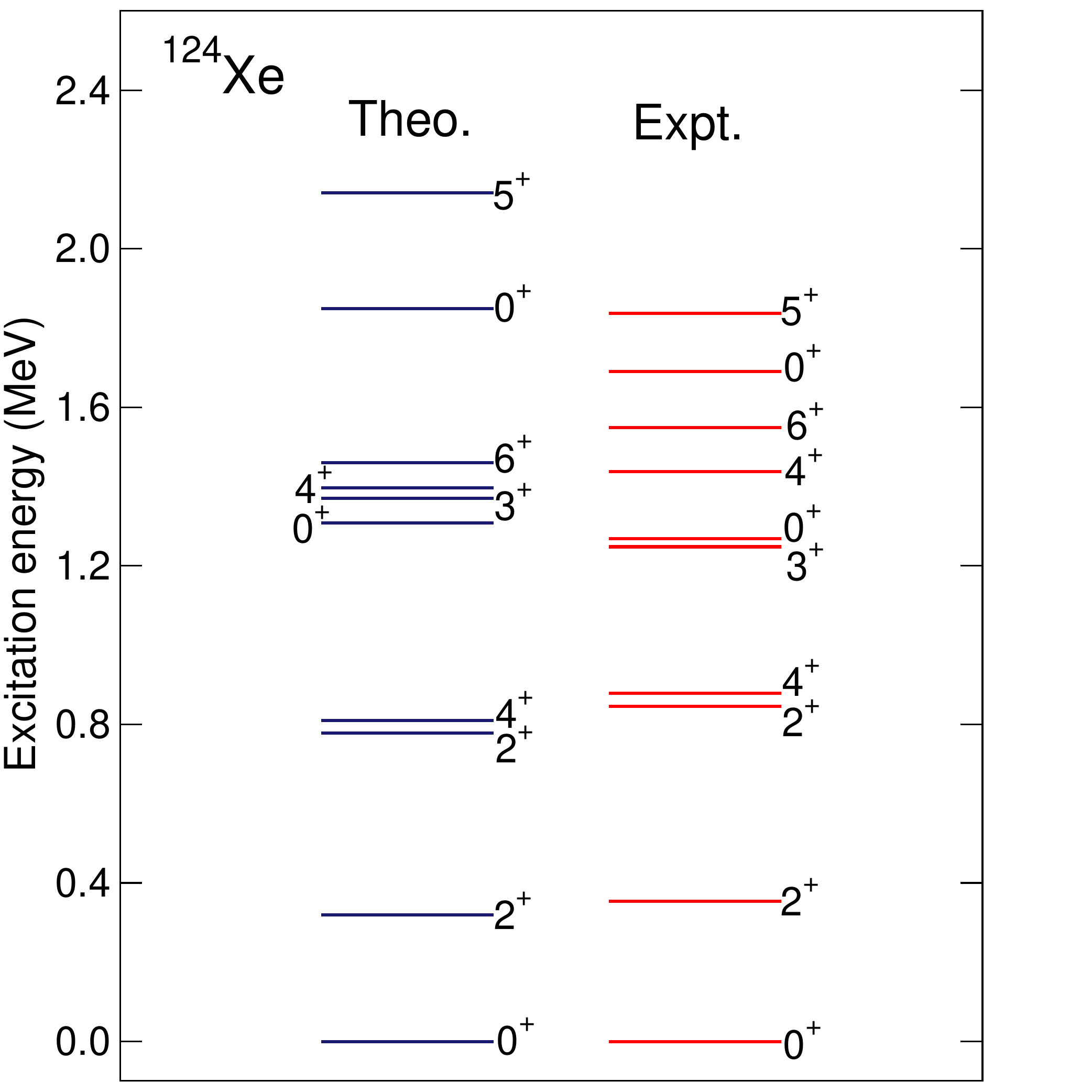} 
\caption{(Color online) Theoretical and experimental \cite{data} 
low-energy excitation spectra for 
 $^{124}$Xe.}
\label{fig:124xe-level} 
\end{center}
\end{figure}

In Ref.~\cite{nomura2017odd-3}, we have considered transitions from 
$\gamma$-soft to nearly spherical shapes in the even-even isotopes 
$^{126-136}$Xe as well as in the case of  odd-mass Xe and Cs nuclei. 
The same Gogny-D1M energy surfaces for $^{126-132}$Xe used in that work 
have been used to fix, this time, the parameters of the  IBM-2 
Hamiltonian for these nuclei. Only the energy surface of $^{124}$Xe  
has been added to the results obtained in previous calculations. A 
major difference with respect to Ref.~\cite{nomura2017odd-3} is that 
now we use the IBFM-2 instead of the IBFM-1 model, which does not 
distinguish between neutron and proton bosons. Another minor difference 
with respect to Ref.~\cite{nomura2017odd-3} is that now the even-even 
$^{A+1}$Xe$_{N+1}$ nucleus is taken as a reference to obtain the 
results for the odd-$N$ isotope $^{A}$Xe$_{N}$.

The Gogny-D1M and the (mapped) IBM-2 energy surfaces obtained for  
$^{124}$Xe are depicted in Fig.~\ref{fig:124xe-pes}. The HFB energy 
surface exhibits a shallow triaxial minimum with $\gamma\approx 
30^{\circ}$. Such a triaxial minimum can only be obtained in the IBM-2 
after including higher-order (e.g., three-body) terms. We are, 
however, neglecting such higher-order terms in this study because of 
the lack of  IBFFM and IBFM computer codes able to handle them. As seen 
in Fig.~\ref{fig:124xe-pes}, the IBM-2 surface is much flatter than the 
HFB far away from the global mean-field minimum. This is a consequence 
of the reduced IBM model space and it has already been found and 
discussed in great details in our previous studies 
\cite{nomura2008,nomura2010}. These are not serious limitations as the 
most relevant configurations for the study of low-lying collective 
states are those around the global minimum and we have paid special 
attention to reproduce them.

The energy spectrum provided by the IBM-2 Hamiltonian for $^{124}$Xe  
is compared  in Fig.~\ref{fig:124xe-level} with the experimental data 
\cite{data}. As can be seen, our calculations reproduce well the 
experimental spectrum without any phenomenological adjustment. Both the 
theoretical and experimental spectra exhibit features resembling those 
of the O(6) dynamical symmetry, i.e., $R_{4/2}=E(4^+_1)/E(2^+_1) 
\approx 2.5$, a low-lying  $2^+_2$ level close to the $4^+_1$ one and 
the nearly staggered  energy systematic of the $\gamma$-band (i.e., 
$2^+_2$, ($3^+_1$, $4^+_2$), ($5^+_1$, $6^+_2$), $\ldots$ etc).


\subsection{Odd-mass nuclei\label{sec:oe}}


\begin{figure*}[htb!]
\begin{center}
\includegraphics[width=0.6\linewidth]{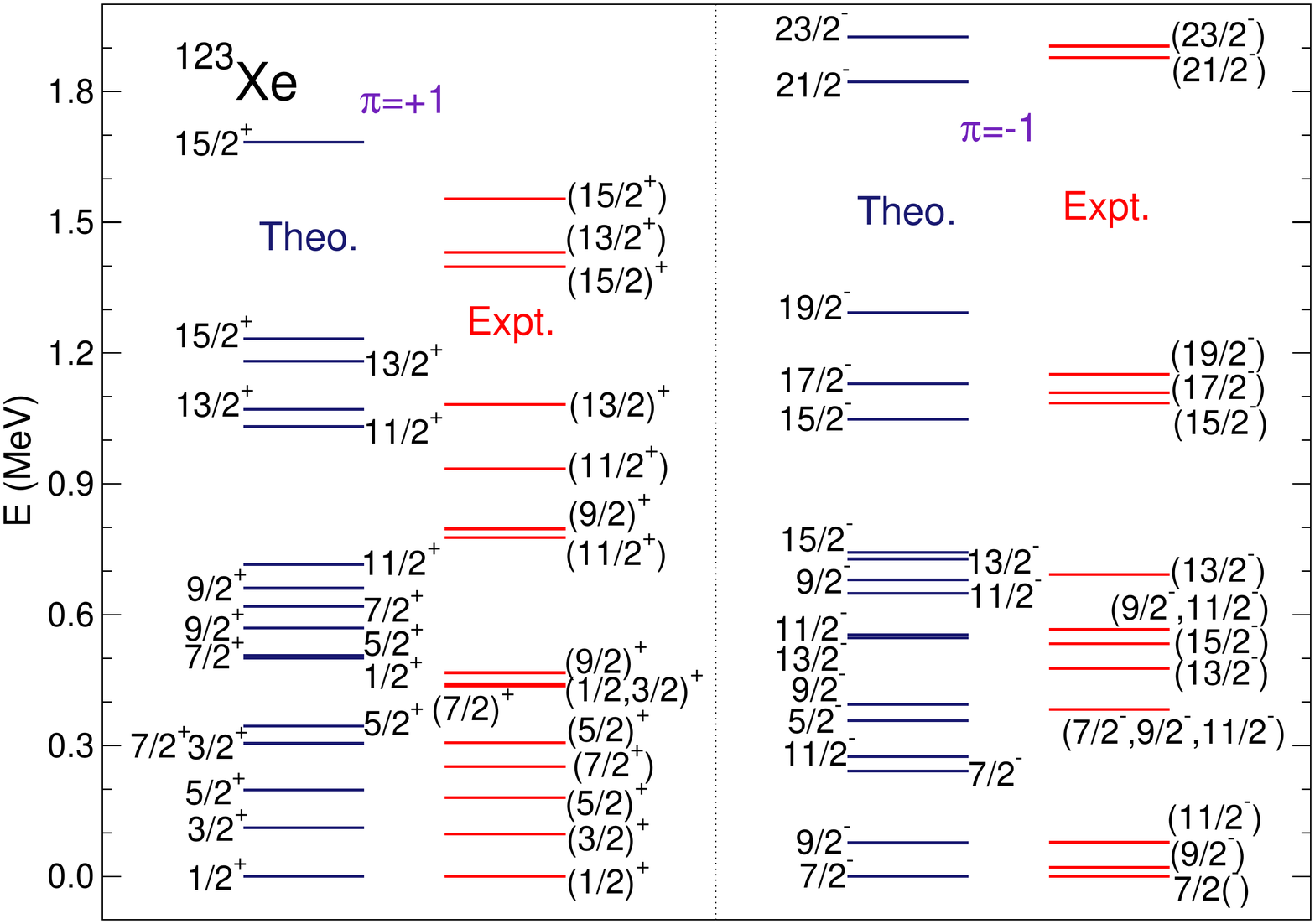}\\
\includegraphics[width=0.6\linewidth]{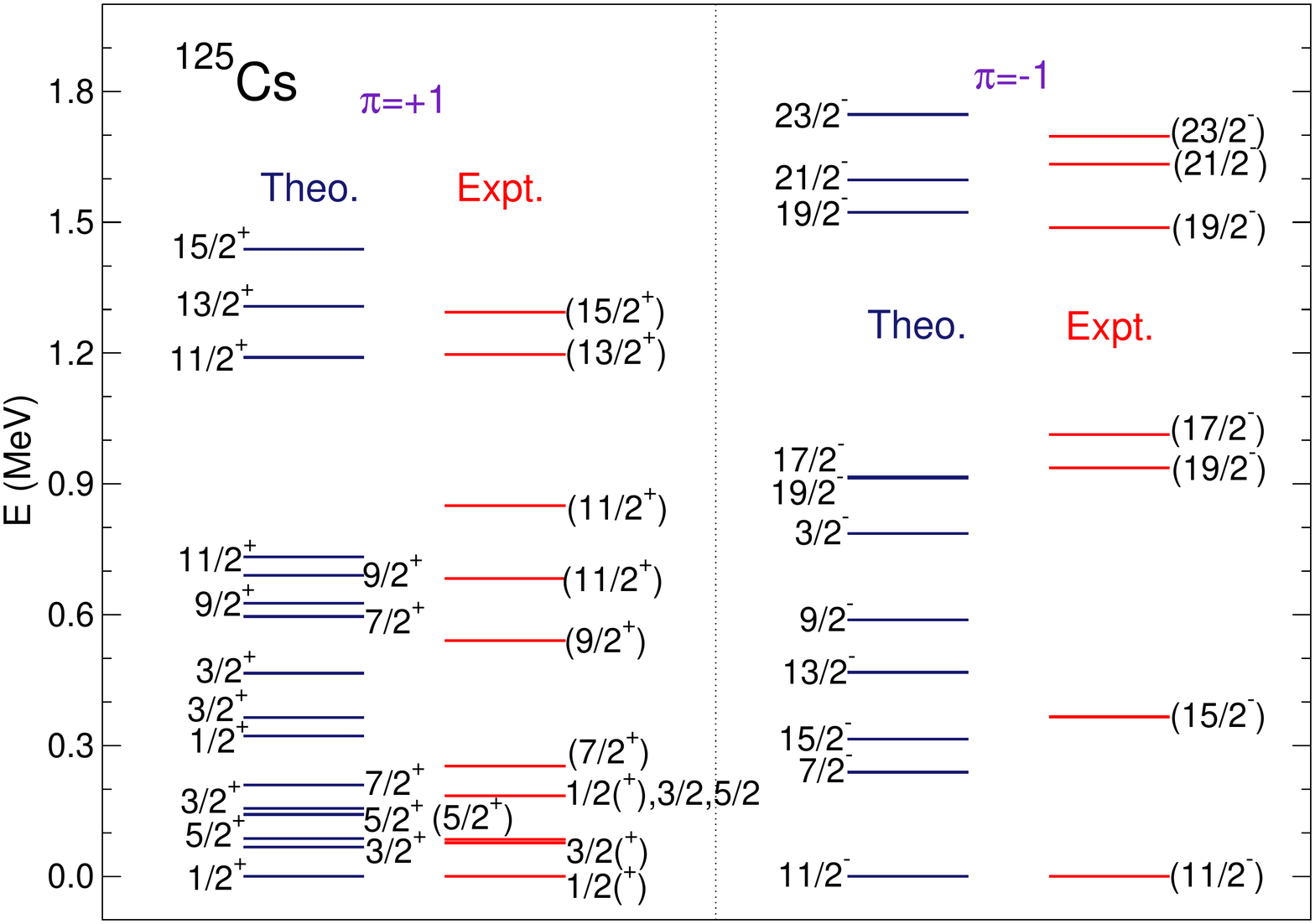}
\caption{Same as Fig.~\ref{fig:124xe-level}, but for the odd-N 
 $^{123}$Xe and odd-Z  $^{125}$Cs nuclei. The spin and/or parity in 
 parentheses have not been  established experimentally. }
\label{fig:oe-level} 
\end{center}
\end{figure*}

Let us turn our attention to the nuclei $^{123}$Xe and $^{125}$Cs. The 
low-lying positive- and negative-parity states obtained for those 
nuclei are shown in Fig.~\ref{fig:oe-level}. They are compared with the 
available experimental data \cite{data}. Our results suggest that the 
low-lying positive-parity states in $^{123}$Xe are mainly built via the 
coupling of the odd neutron hole in the $3s_{1/2}$ and $2d_{3/2}$ 
single-particle orbitals to the even-even boson core ($^{124}$Xe). On 
the other hand, the negative-parity states are accounted for  by the 
unique-parity $1h_{11/2}$ single-particle configuration. As seen in 
Fig.~\ref{fig:oe-level} our results agree well with the experiment for 
both parities. In the case of  $^{125}$Cs, the low-lying 
positive-parity states are mainly based on the $1g_{7/2}$ and 
$2d_{5/2}$ single-particle configurations.  In the lower panels of 
Fig.~\ref{fig:oe-level} a reasonable agreement between the predicted 
IBFM-2 and the experimental spectra is observed.

\subsection{Odd-odd Cs isotopes\label{sec:oo}}


\begin{figure*}[htb!]
\begin{center}
\includegraphics[width=0.6\linewidth]{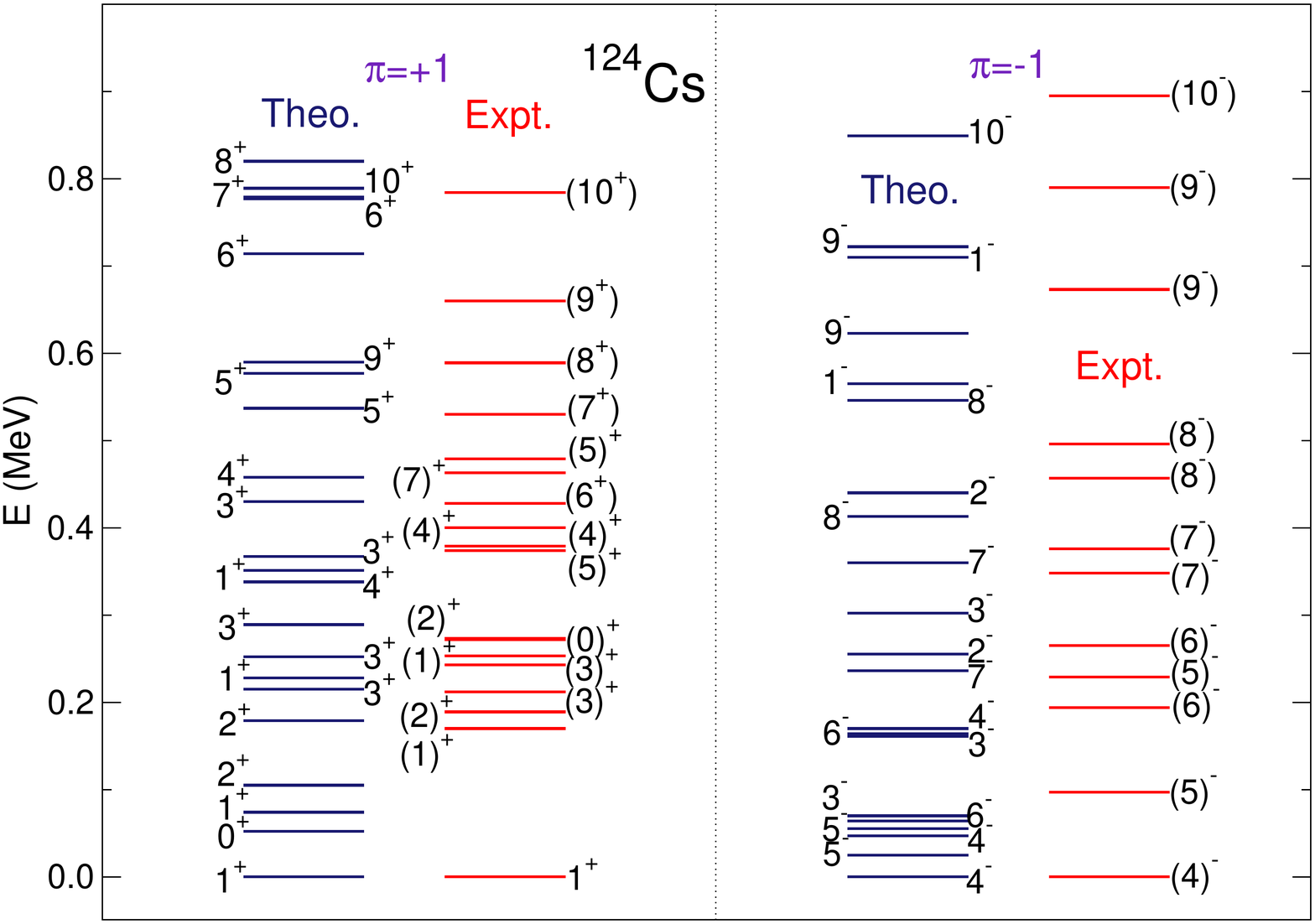}
\caption{(Color online) Low-lying positive-
and negative-states of the odd-odd nucleus $^{124}$Cs. 
Experimental energy levels are taken from Ref.~\cite{data}. }
\label{fig:124cs} 
\end{center}
\end{figure*}

\begin{figure*}[htb!]
\begin{center}
\includegraphics[width=0.6\linewidth]{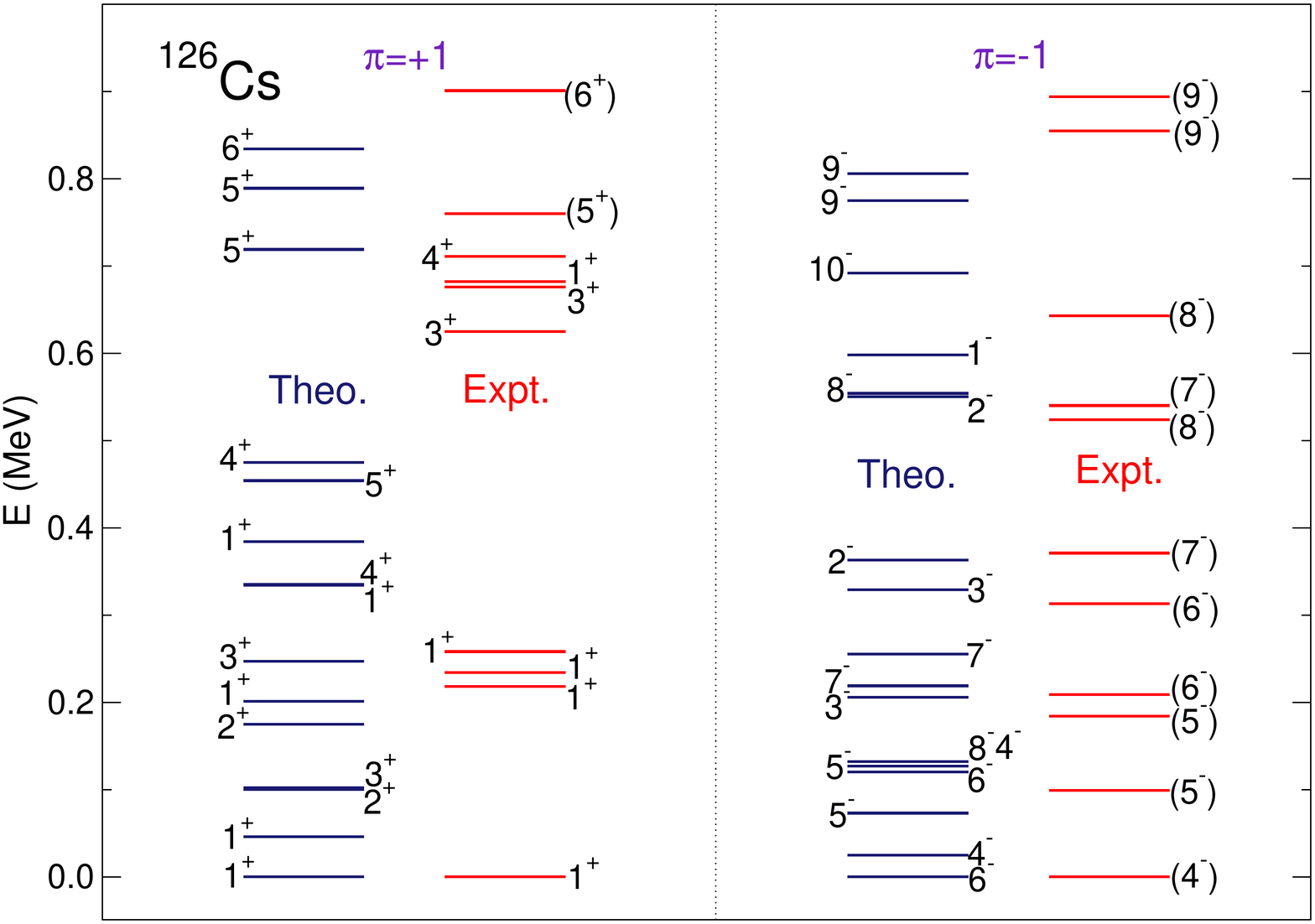}
\caption{(Color online) Same as in Fig.~\ref{fig:124cs} but for 
$^{126}$Cs. Experimental data for  positive- and 
negative-parity states are taken from Refs.~\cite{data} and 
\cite{li2003cs126}, respectively.}
\label{fig:126cs} 
\end{center}
\end{figure*}

\begin{figure*}[htb!]
\begin{center}
\includegraphics[width=0.6\linewidth]{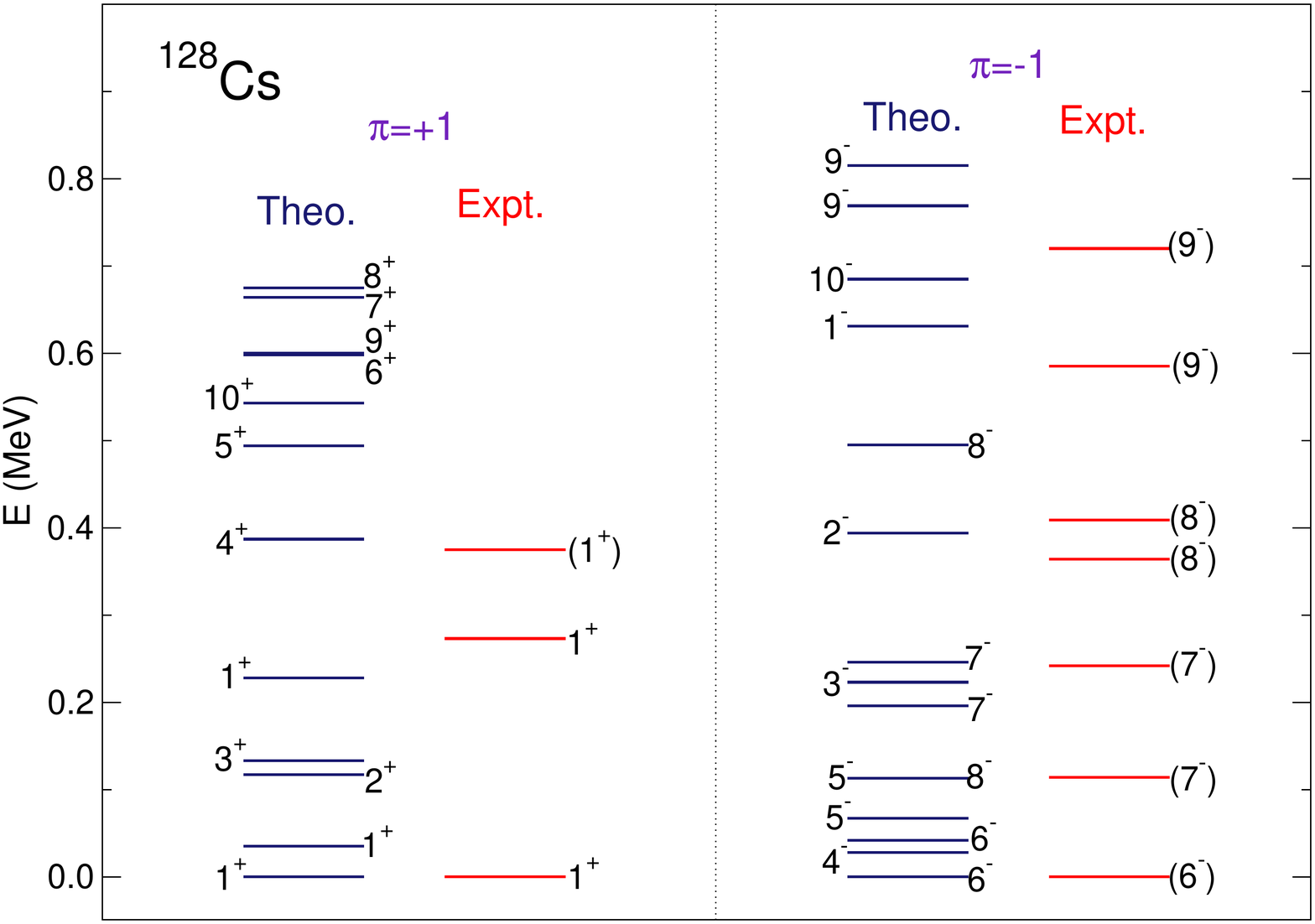}
\caption{Same as in Fig.~\ref{fig:124cs} but for 
$^{128}$Cs.}
\label{fig:128cs} 
\end{center}
\end{figure*}

\begin{figure*}[htb!]
\begin{center}
\includegraphics[width=0.6\linewidth]{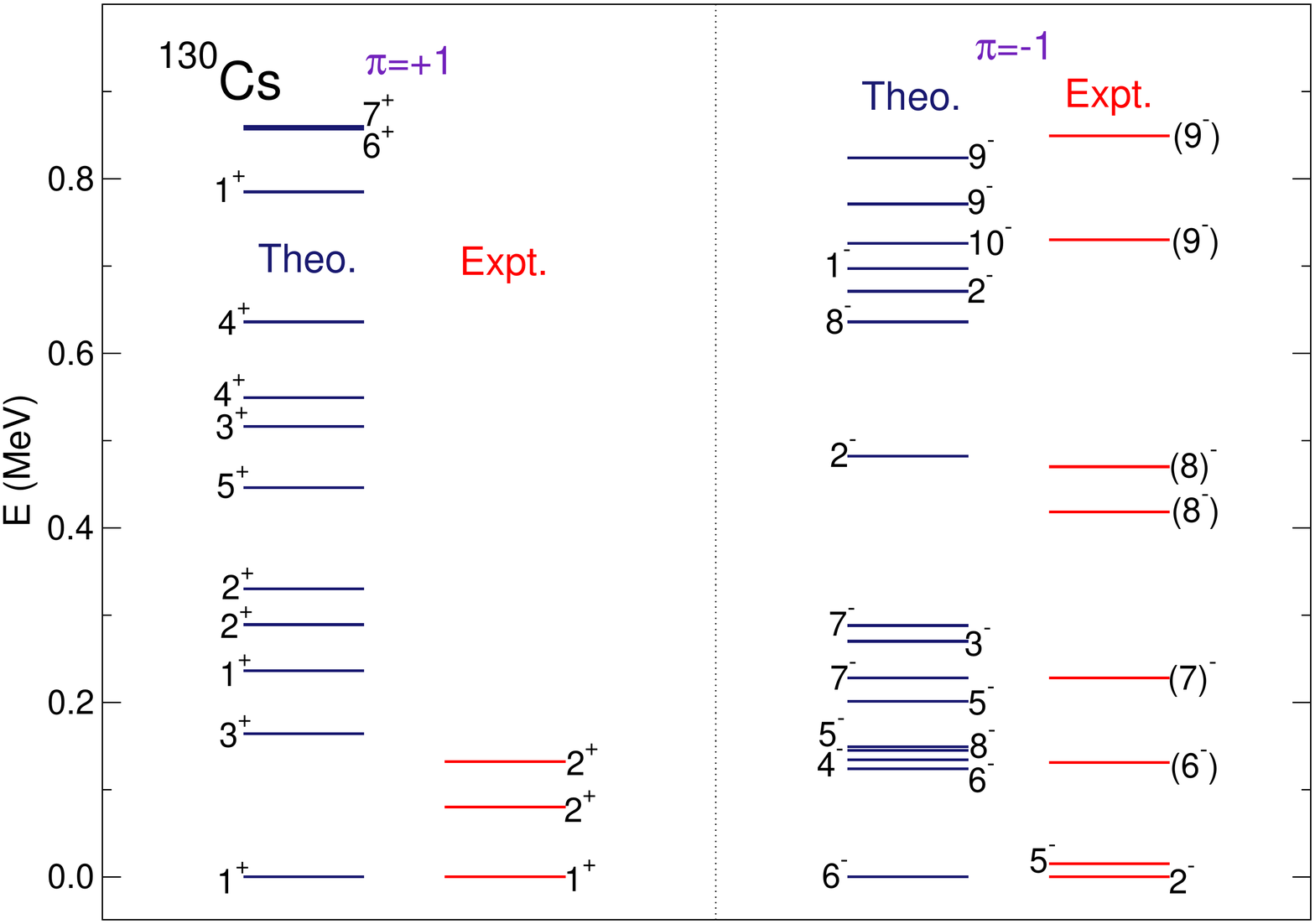}
\caption{Same as in Fig.~\ref{fig:124cs} but for 
$^{130}$Cs.}
\label{fig:130cs} 
\end{center}
\end{figure*}

\begin{figure*}[htb!]
\begin{center}
\includegraphics[width=0.6\linewidth]{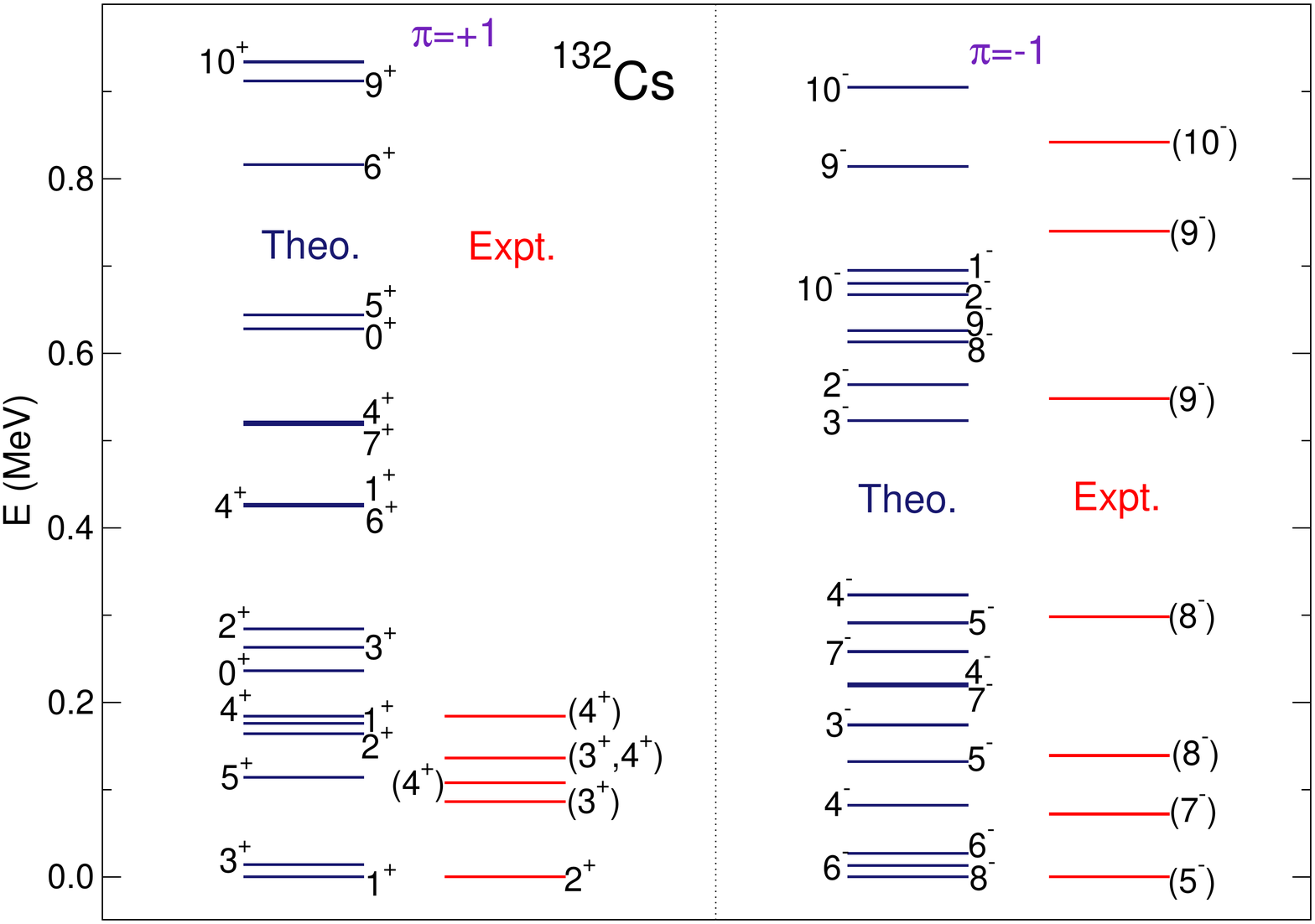}
\caption{Same as in Fig.~\ref{fig:124cs} but for 
$^{132}$Cs.}
\label{fig:132cs} 
\end{center}
\end{figure*}


\subsubsection{Energy spectra for the low-spin low-energy states}


Let us now discuss the results obtained for odd-odd Cs nuclei. We will 
consider  low-spin low-energy states up to an excitation energy $E_{\rm 
x} \approx 1$ MeV. Our calculation indicates that those states are 
mainly based on normal-parity (i.e., $sdg$) orbitals.

The spectra obtained for  $^{124,126,128,130,132}$Cs are depicted in 
Figs.~\ref{fig:124cs}--\ref{fig:132cs}, respectively. In the case of 
$^{124}$Cs  (see, Fig.~\ref{fig:124cs}), the predicted positive- and 
negative-parity states agree well with the experimental ones. The 
IBFFM-2 wave function of the $1^+_1$ ground state is composed of the 
mixture of several  single-particle configurations among which, the 
largest (about 50 \%) contribution comes from the  odd neutron hole in 
the $3s_{1/2}$ orbital. As for the negative-parity states, the 
predicted IBFFM-2 wave functions for the lowest  $4^-_1$, $5^-_1$, and 
$6^-_1$ states are complex mixtures of different single-particle 
configurations. In those states, the neutron $\nu h_{11/2}$ coupled to 
the proton in either $3s_{1/2}$, $2d_{3/2}$, $2d_{5/2}$, or $1g_{7/2}$ 
positive-parity orbital plays a dominant role. 

In the case of $^{126}$Cs (see, Fig.~\ref{fig:126cs}), the agreement 
with the experiment is as good as for $^{124}$Cs. The structure of the 
wave functions corresponding to the lowest positive-parity states is 
similar to the one obtained for $^{124}$Cs (i.e., they are mainly 
accounted for by the $(\nu s_{1/2})^{-1}\otimes\pi sdg$ configuration). 
In our calculation the lowest-energy negative-parity state is predicted 
to be the $6^-_1$ one. The main component (47 \%) of the IBFFM-2 wave 
function of this $6^-_1$ state is the configuration 
$[(\nu h_{11/2})^{-1}\otimes\pi g_{7/2}]^{(J=8^-)}$. Experimentally, the 
$4^-$ state is suggested to be the lowest negative-parity state, and 
the tentative $6^-_1$ level is found at a much higher excitation energy than in 
our calculation. However, for most of the low-lying negative-parity 
states both spin and parity have not been firmly established. 

The experimental data are more scarce for the $^{128}$Cs nucleus, as 
well as for the heavier ones $^{130,132}$Cs. For the nucleus 
$^{128}$Cs, experimental information is only available for a couple of 
$1^{+}$ states. Here, we stress that our calculations reproduce the 
correct ground-state spin $I=1^{+}_{1}$. Note, that the predicted 
$1^+_2$ and $1^+_3$ non-yrast states are found below 200 keV excitation 
energy, somewhat similar to the experimental situation. Furthermore, we 
also obtain $2^+$ and $3^+$ states below 200 keV. The structure of the 
$1^+_1$, $2^+_1$, and $3^+_1$ wave functions is similar to the one in 
$^{124}$Cs and $^{126}$Cs. Concerning the negative-parity states of 
$^{128}$Cs, the predicted low-spin  levels are in reasonable agreement with the 
experimental ones. However, our calculations suggest several states 
near the ground state, that have not been observed experimentally 
(i.e., a $4^-$ and two $5^-$ states, and the second $6^-$ state). 

The positive-parity low-spin spectrum  obtained for $^{130}$Cs is shown 
in Fig.~\ref{fig:130cs}. Once more, our calculations predict the 
correct ground-state spin  $I=1^+$. However, the two experimental $2^+$ 
states around 100 keV excitation energy are overestimated by a factor 
of three. This is not surprising as the excitation energy of levels is 
often overestimated within the IBM framework, for those nuclei near a 
shell closure and the reason is the decreasing number of active bosons. 
This also seems to be the case for both $^{130}$Cs and $^{132}$Cs. In 
addition, the structure of the  IBFFM-2 wave function corresponding to 
the $1^+_1$ state turns out to be slightly different than the ground 
states of the lighter odd-odd systems $^{124-128}$Cs. The contribution 
of the $\nu d_{3/2}$ single-particle configuration becomes larger  in 
$^{130,132}$Cs than in $^{124-128}$Cs. The HFB deformation energy 
surfaces obtained for even-even Xe isotopes \cite{nomura2017odd-3} 
exhibit a structural change from $^{128}$Xe ($\gamma$-soft shape with a 
shallow triaxial minimum) to $^{130}$Xe (nearly spherical shape  with a 
shallow prolate minimum). Such a  structural change in the even-even 
systems seems to be more or less translated into the structure of the 
IBFFM-2 wave functions of the odd-odd systems. 
As can be seen from Fig.~\ref{fig:130cs}, the disagreement with the 
experimental data is more pronounced for negative-parity states. The 
energies of the $2^-_1$ and $5^-_1$ levels, which are suggested to be 
the lowest negative-parity states experimentally, are however too high 
in our calculations. 

Finally, the positive- and negative-parity low-spin low-energy spectra 
obtained for $^{132}$Cs are depicted in Fig.~\ref{fig:132cs}. Here, the 
comparison with the experiment is worse, but one should keep in mind 
that this nucleus is the closest to the $N=82$  shell closure. As a 
result, the number of  neutron $N_{\nu}=2$ and proton $N_{\pi}=2$ 
bosons is probably not enough for a detailed description of the level 
structure in the framework of the IBM. Other possible reasons are first 
that the single-particle energies and occupation probabilities for odd 
nucleons, obtained from the Gogny-D1M calculation, may not be realistic 
enough in this case. Finally, the fixed values of the strengths and/or 
the  forms of the residual neutron-proton interactions employed  in the 
IBFFM-2 Hamiltonian are too restrictive.


\subsubsection{E2 and M1 moments of lowest-lying states}


\begin{table}[!htb]
\begin{center}
\caption{\label{tab:oo-mom} Theoretical and experimental quadrupole $Q(I)$
 (in $e$b units) and magnetic $\mu(I)$ (in $\mu_N$ units) moments for
 $^{124-132}$Cs. The experimental values are taken
 from Ref.~\cite{stone2005}.}
 \begin{ruledtabular}
 \begin{tabular}{cccc}
  & & Theory & Experiment \\
\hline
$^{124}$Cs & $Q(1^+_1)$ & $-$0.475 & $-$0.74(3) \\
           & $\mu(1^+_1)$ & $+$0.377 & $+$0.673(3) \\
$^{126}$Cs & $Q(1^+_1)$ & $-$0.585 & $-$0.68(2) \\
           & $\mu(1^+_1)$ & $+$0.869 & $+$0.777(4) \\
$^{128}$Cs & $Q(1^+_1)$ & $-$0.471 & $-$0.570(8) \\
           & $\mu(1^+_1)$ & $+$0.794 & $+$0.974(5) \\
$^{130}$Cs & $Q(1^+_1)$ & $-$0.125 & $-$0.059(6) \\
           & $\mu(1^+_1)$ & $+$0.573 & $+$1.460(7) \\
           & $Q(5^-_1)$ & $-$0.314 & $+$1.45(5) \\
           & $\mu(5^-_1)$ & $-$1.062 & $+$0.629(4) \\
$^{132}$Cs & $Q(2^+_1)$ & $-0.062$ & $+$0.508(7) \\
           & $\mu(2^+_1)$ & $+0.940$ & $+$2.222(7)
 \end{tabular}
 \end{ruledtabular}
\end{center} 
\end{table}

As for the electromagnetic properties of the lowest-lying states in 
odd-odd Cs isotopes, experimental data are only available  for the 
quadrupole $Q(I)$ and magnetic dipole $\mu(I)$ moments. The theoretical 
and the available experimental $Q(I)$ and $\mu(I)$ values are compared 
in Table~\ref{tab:oo-mom}. For the $^{124,126,128}$Cs nuclei, the 
predicted $Q(I)$ and $\mu(I)$ moments agree well with the experimental 
ones, in both magnitude and sign. However, some of the moments obtained 
for some states in $^{130,132}$Cs are opposite in sign to their 
experimental counterparts. This corroborates that the 
energy levels of the corresponding states in these nuclei have 
not been described well with respect to the experimental data 
(see, Figs.~\ref{fig:130cs} and \ref{fig:132cs}), and could have 
occurred because of the assumption of using the fixed strength 
parameters for the residual neutron-proton interaction 
$\hat V_\mathrm{res}$ and/or, again, because of the more restricted 
configuration space for the boson system. 

\begin{figure*}[htb!]
\begin{center}
\includegraphics[width=0.6\linewidth]{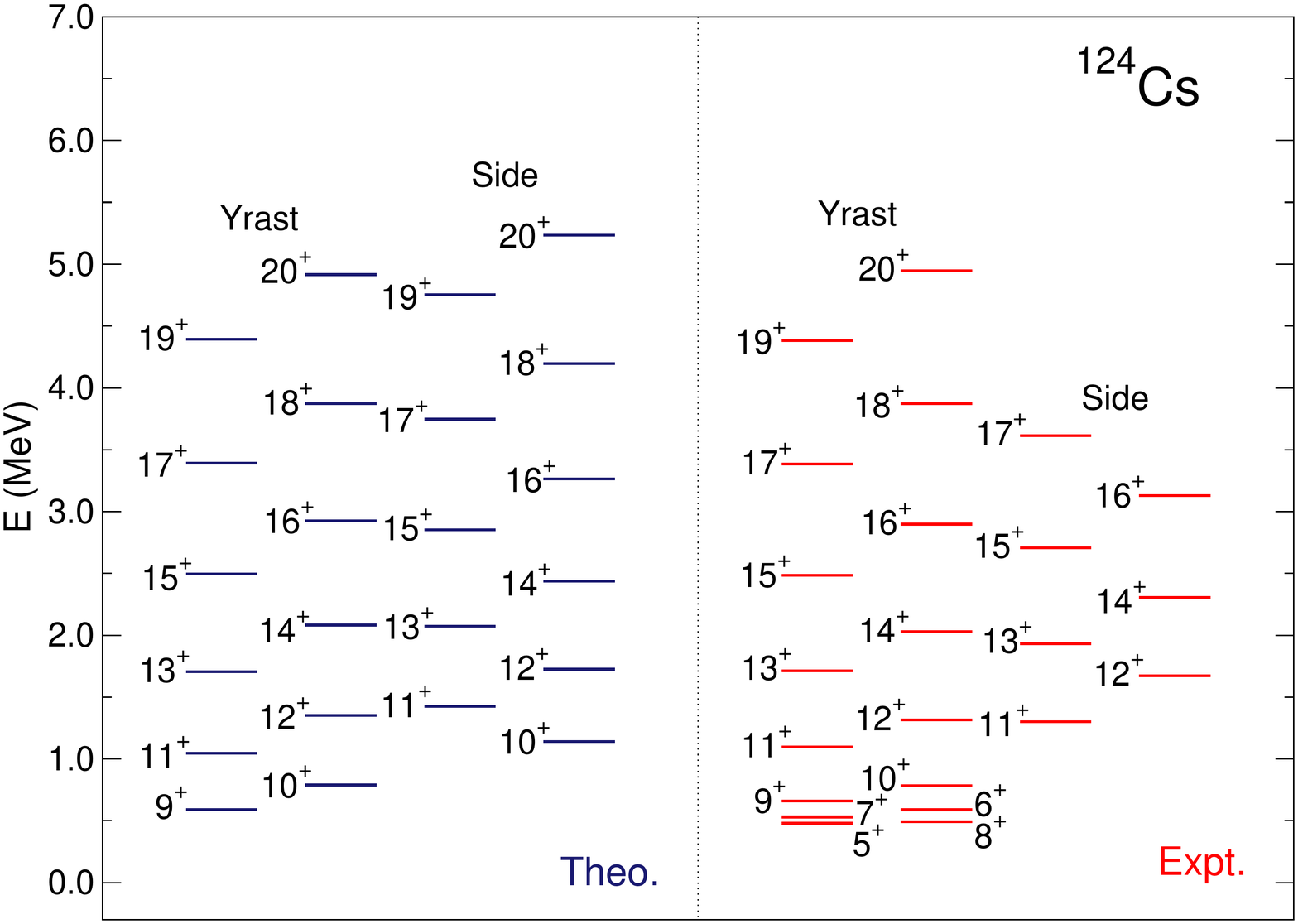}
\caption{(Color online) Band structure of the higher-spin
 higher-energy positive-parity states in $^{124}$Cs.}
\label{fig:124cs-hs} 
\end{center}
\end{figure*}

\begin{figure*}[htb!]
\begin{center}
\includegraphics[width=0.6\linewidth]{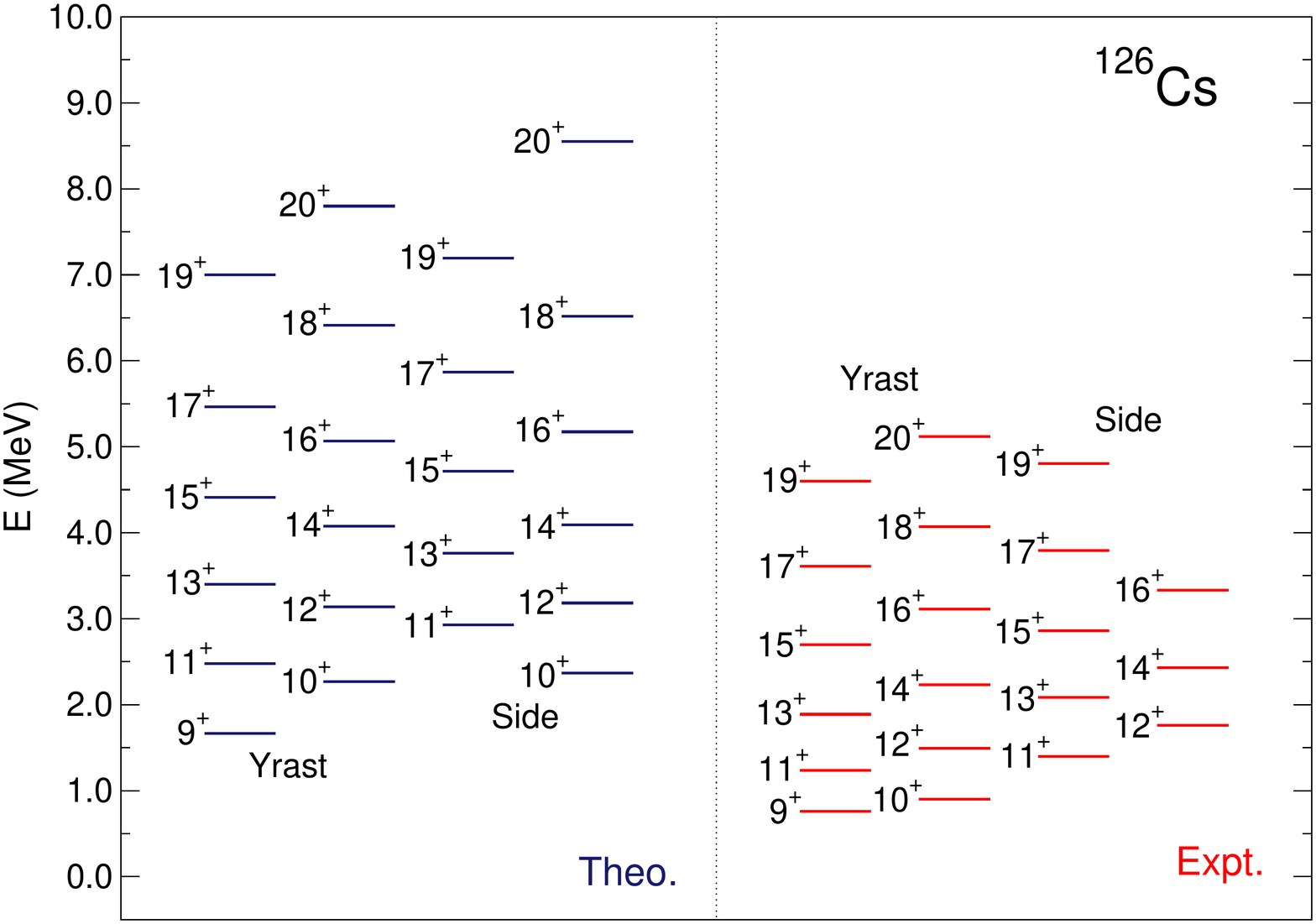}
\caption{(Color online) Same as in Fig.~\ref{fig:124cs-hs} but for  $^{126}$Cs.}
\label{fig:126cs-hs} 
\end{center}
\end{figure*}

\begin{figure*}[htb!]
\begin{center}
\includegraphics[width=0.6\linewidth]{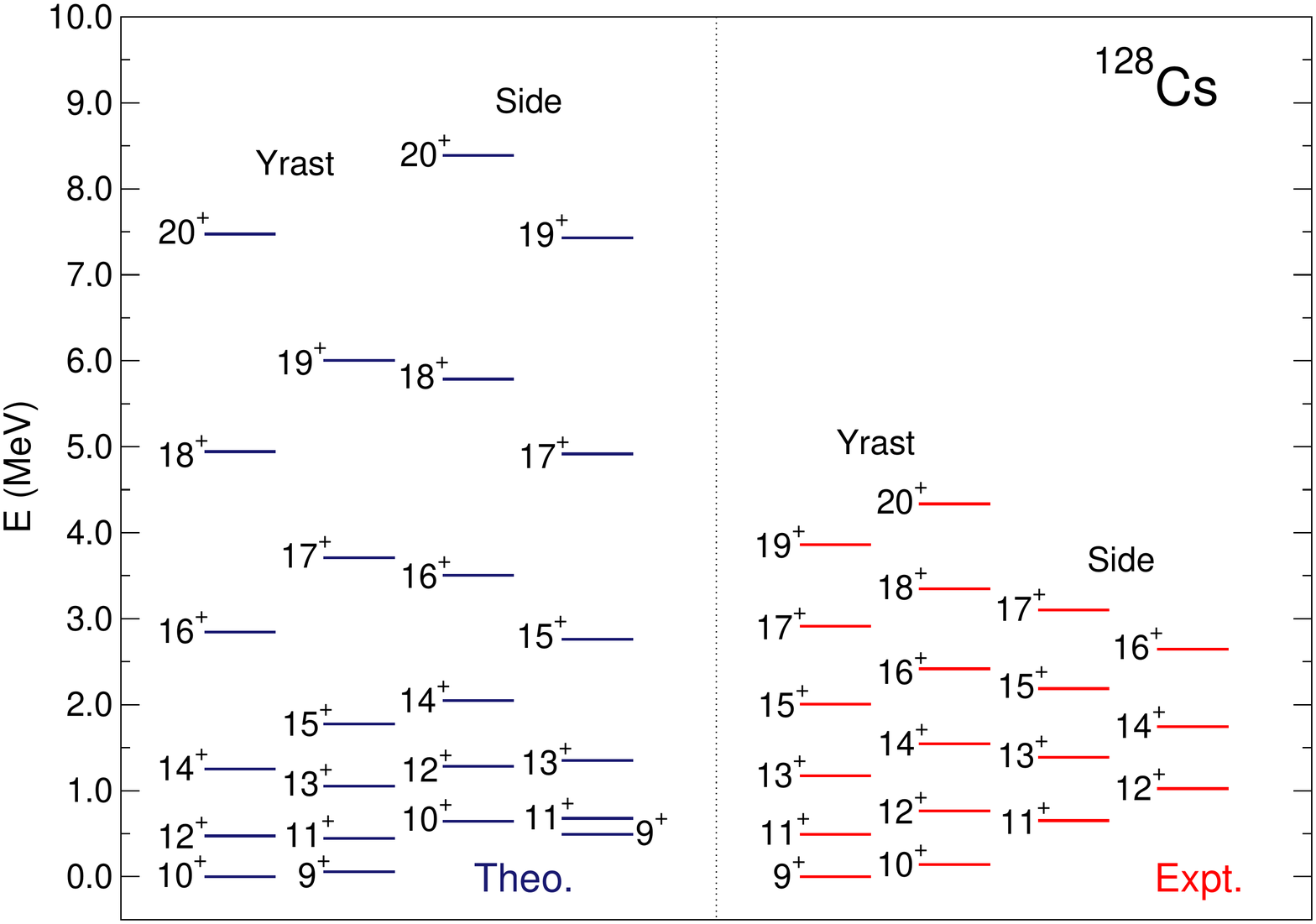}
\caption{(Color online) Same as in Fig.~\ref{fig:124cs-hs} but for $^{128}$Cs. 
The theoretical and experimental spectra are normalized with respect to the 
$10^+_1$ and $9^+_1$ states, respectively, which are the lowest states 
based on the $(\nu h_{11/2})^{-1}\otimes\pi h_{11/2}$ configuration.}
\label{fig:128cs-hs} 
\end{center}
\end{figure*}

\begin{figure*}[htb!]
\begin{center}
\includegraphics[width=0.6\linewidth]{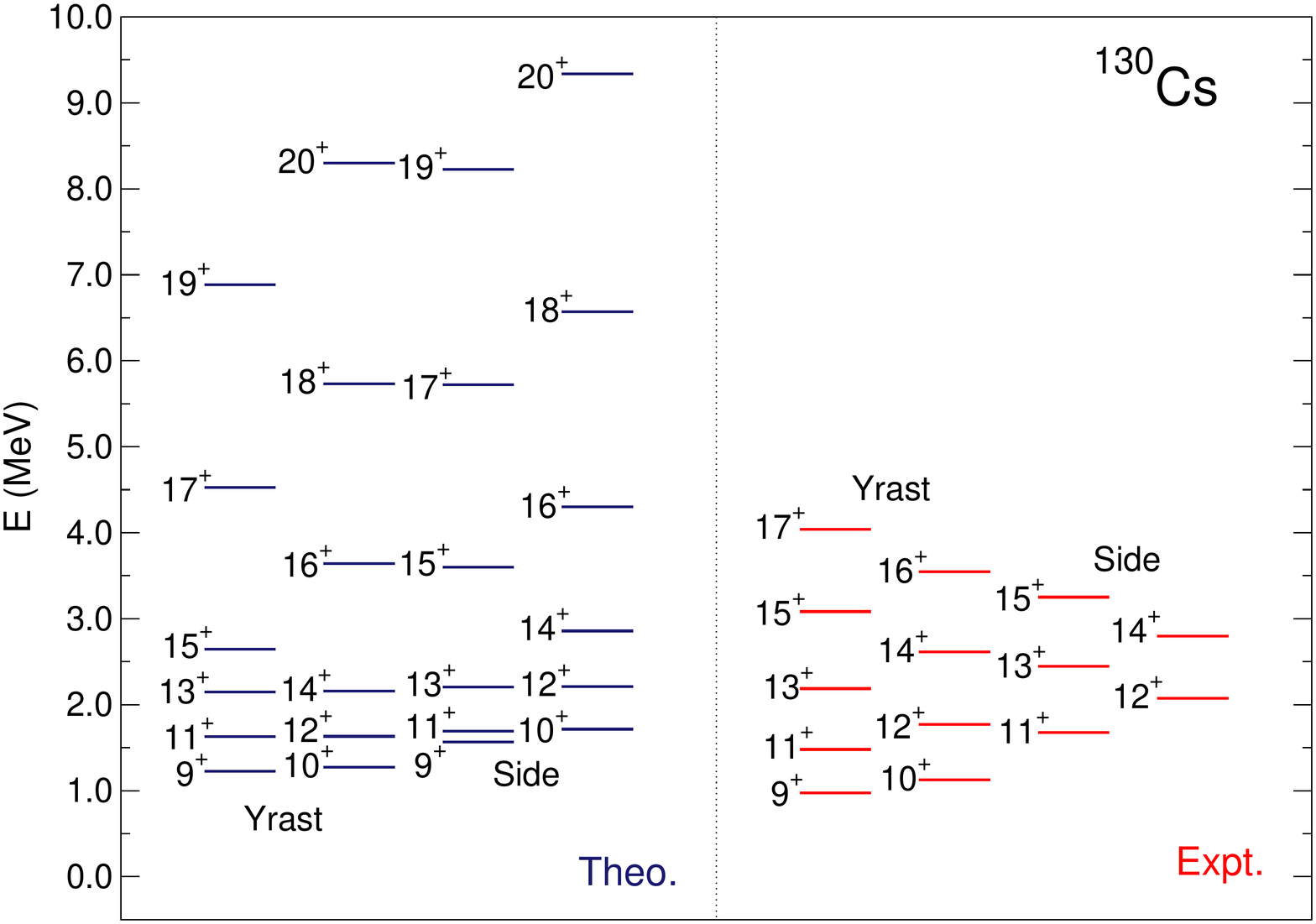}
\caption{(Color online) Same as in Fig.~\ref{fig:124cs-hs} but for  $^{130}$Cs.}
\label{fig:130cs-hs} 
\end{center}
\end{figure*}

\begin{figure*}[htb!]
\begin{center}
\includegraphics[width=0.6\linewidth]{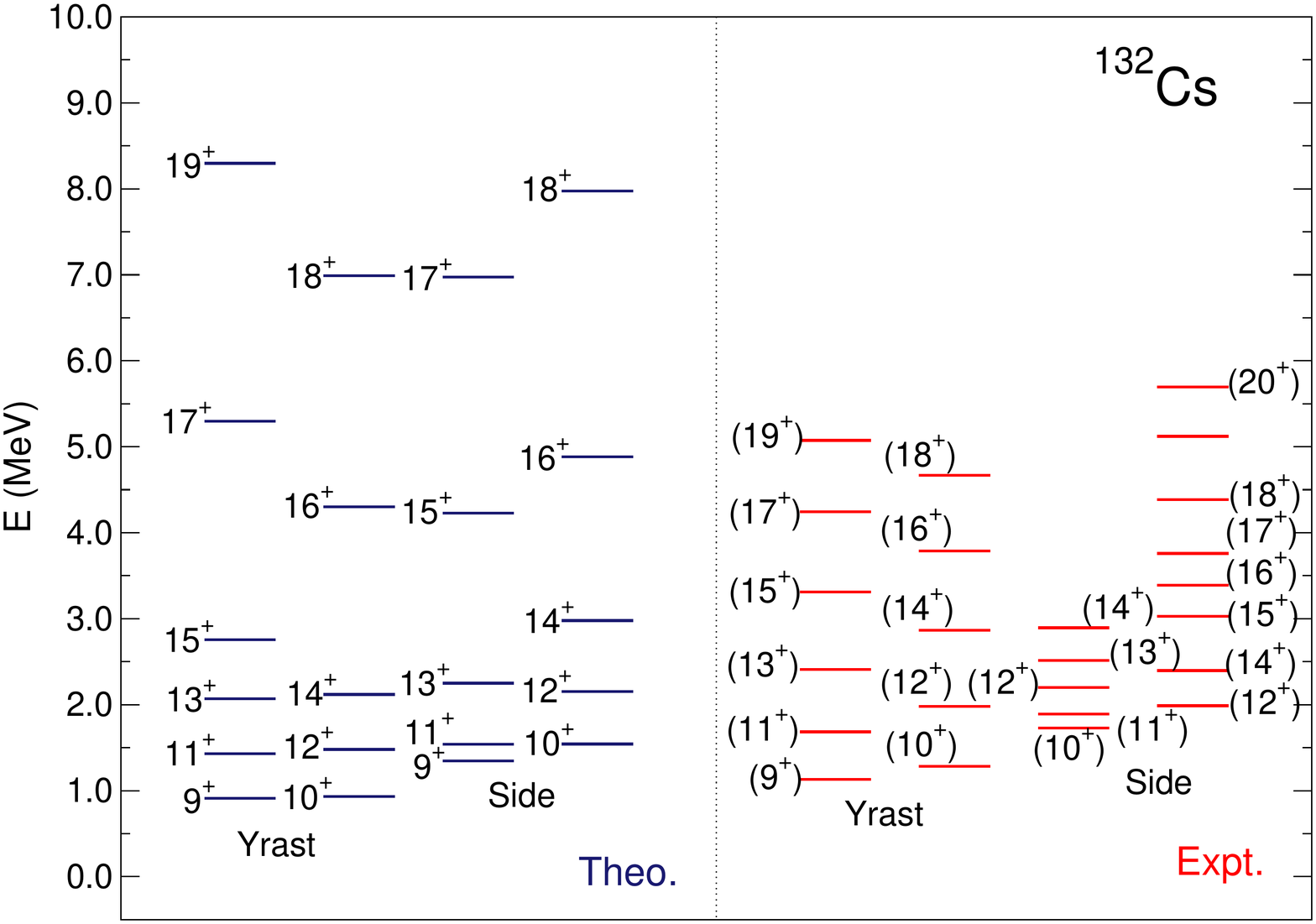}
\caption{(Color online) Same as in Fig.~\ref{fig:124cs-hs} but for  $^{132}$Cs. For the experimental level between the $18^+$ and $20^+$ states in the side band, even the tentative spin and parity are not known \cite{data}.}
\label{fig:132cs-hs} 
\end{center}
\end{figure*}

\begin{figure*}[htb!]
\begin{center}
\includegraphics[width=0.8\linewidth]{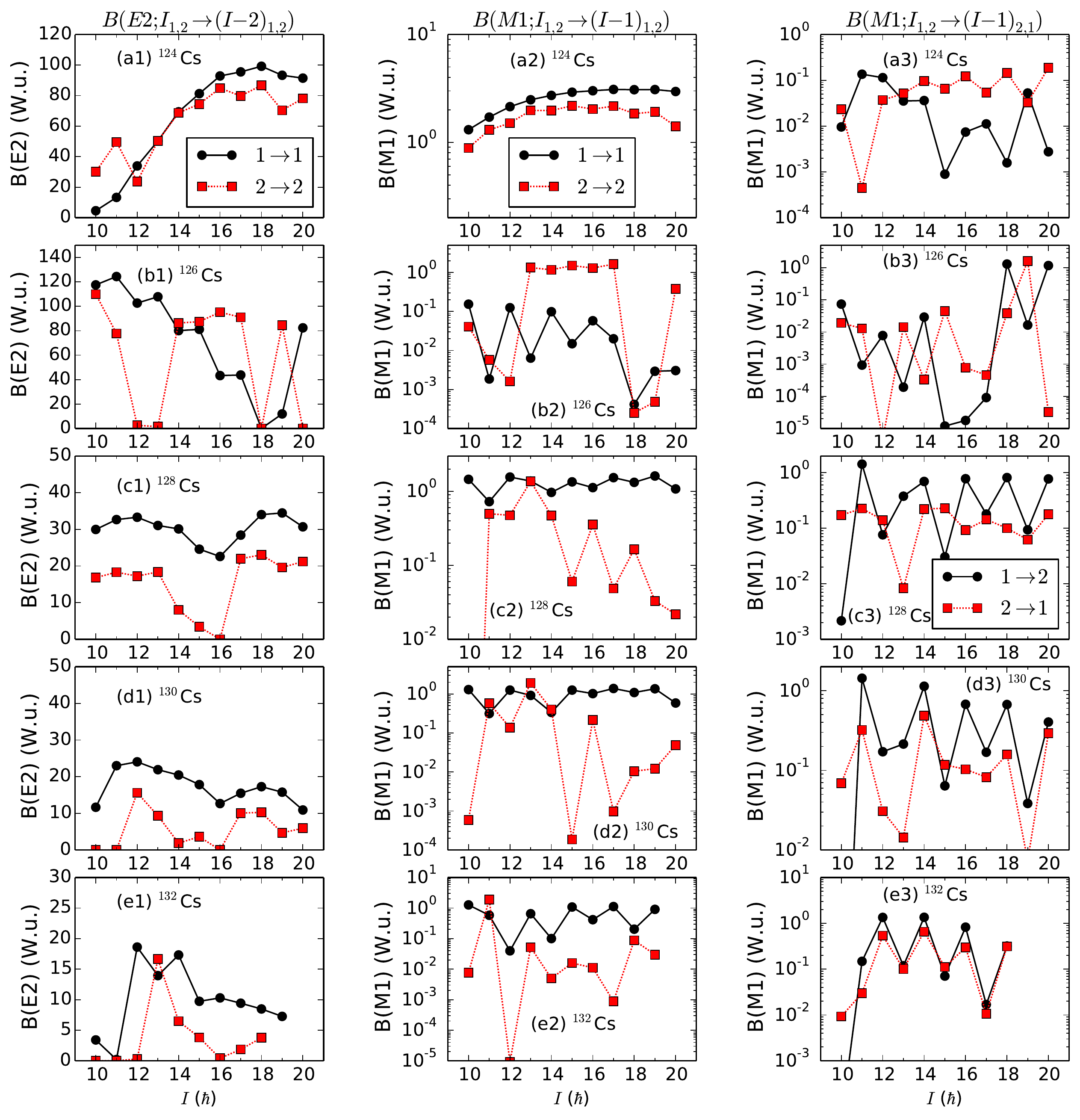}
\caption{(Color online) The calculated $B(E2;I\rightarrow I-2)$ and $B(M1;I\rightarrow I-1)$
 transition strengths (in Weisskopf units) for the positive-parity bands of the $^{124-132}$Cs nuclei. 
Left column (panels (a1) to (e1)): the intra-band $B(E2;I_{1,2}\rightarrow (I-2)_{1,2})$ transition rates 
between the yrast states ($I_1$ and $(I-2)_1$) 
and between the second lowest states ($I_2$ and $(I-2)_2$) with a given spin $I$. 
Middle column (panels (a2) to (e2)): the intra-band $B(M1;I_{1,2}\rightarrow (I-1)_{1,2})$ transition strengths. 
Right column (panels (a3) to (e3)): the inter-band $B(M1;I_{1,2}\rightarrow (I-1)_{2,1})$ transition strengths.}
\label{fig:em-hs} 
\end{center}
\end{figure*}

\begin{figure}[htb!]
\begin{center}
\includegraphics[width=\linewidth]{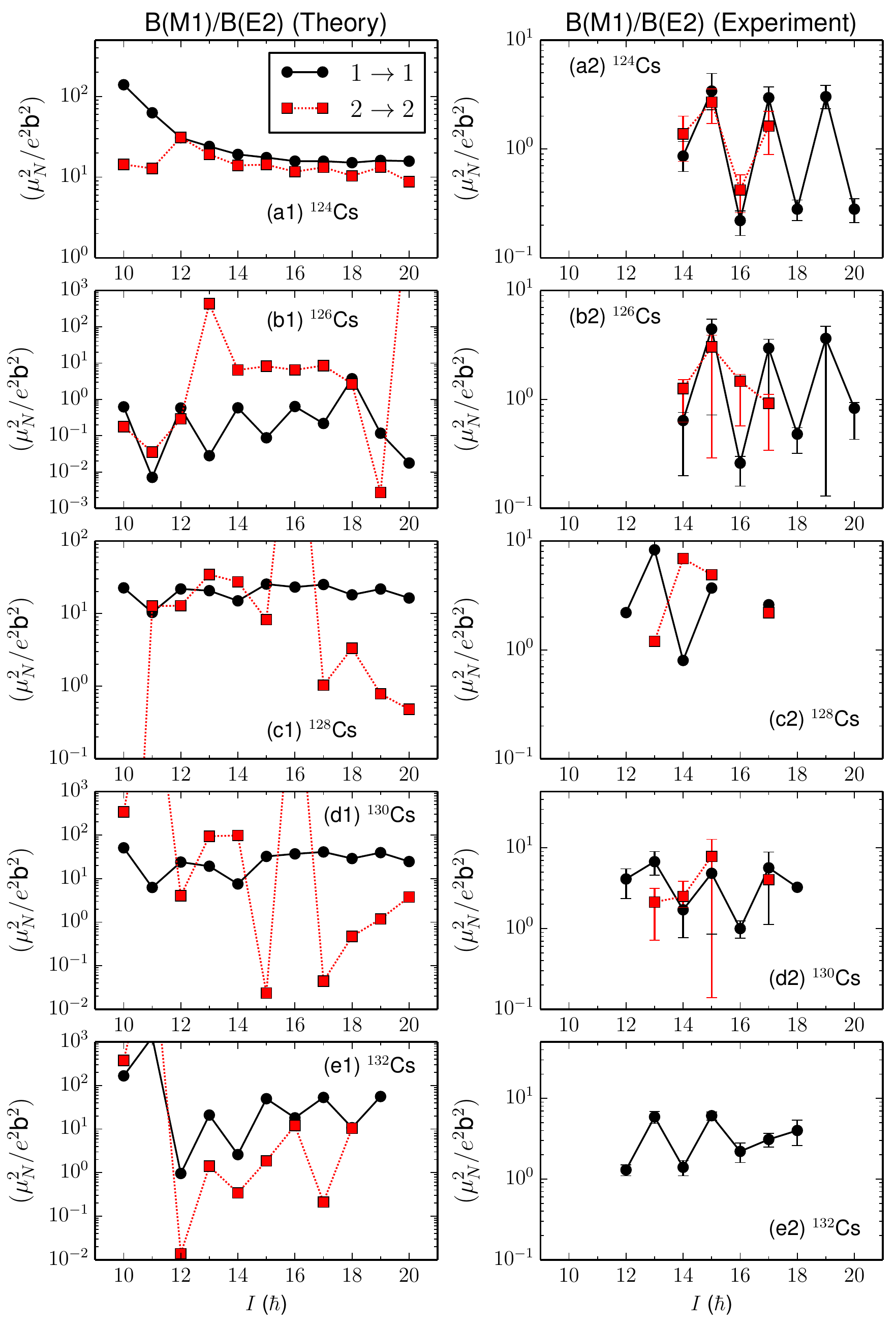}
\caption{(Color online) The calculated and experimental 
values of the ratio $B(M1;I\rightarrow I-1)/B(E2;I\rightarrow I-2)$ (in $\mu_N^2/e^2$b$^2$ units)
are plotted as a function of $I$ for the positive-parity yrast and side bands of the 
odd-odd nuclei $^{124-132}$Cs. The experimental data are 
taken from Refs.~\cite{xiong2019,gizon2001,wang2006,paul1989,simons2005,rainovski2003}. 
Note that calculated $B(M1)/B(E2)$ values lying 
well outside of the scale of the vertical axis 
are not shown. 
}
\label{fig:em-ratio} 
\end{center}
\end{figure}

\begin{figure}[htb!]
\begin{center}
\includegraphics[width=\linewidth]{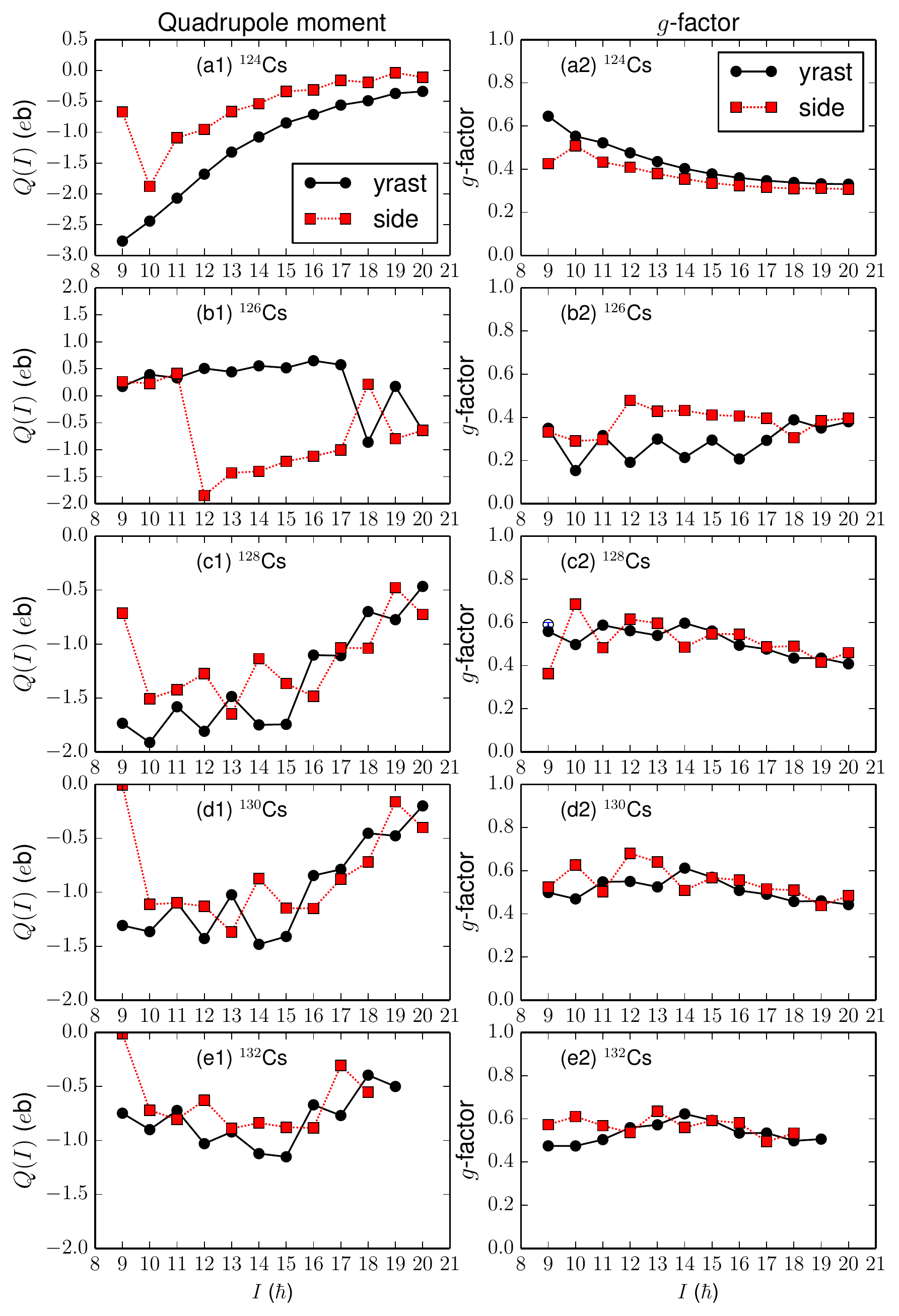}
\caption{(Color online) Electric quadrupole moment $Q(I)$ (in $e$b units) and $g$-factor 
as functions of angular momentum $I$ of the higher-spin yrast and side band states of 
the studied odd-odd Cs isotopes. 
In panel (c2), the experimental value for the $g$-factor of $+0.59\pm 0.01$ 
for the yrast $9^+$ state of $^{128}$Cs \cite{grodner2018} is shown as an open circle.}
\label{fig:mom} 
\end{center}
\end{figure}

\begin{figure*}[htb!]
\begin{center}
\includegraphics[width=0.6\linewidth]{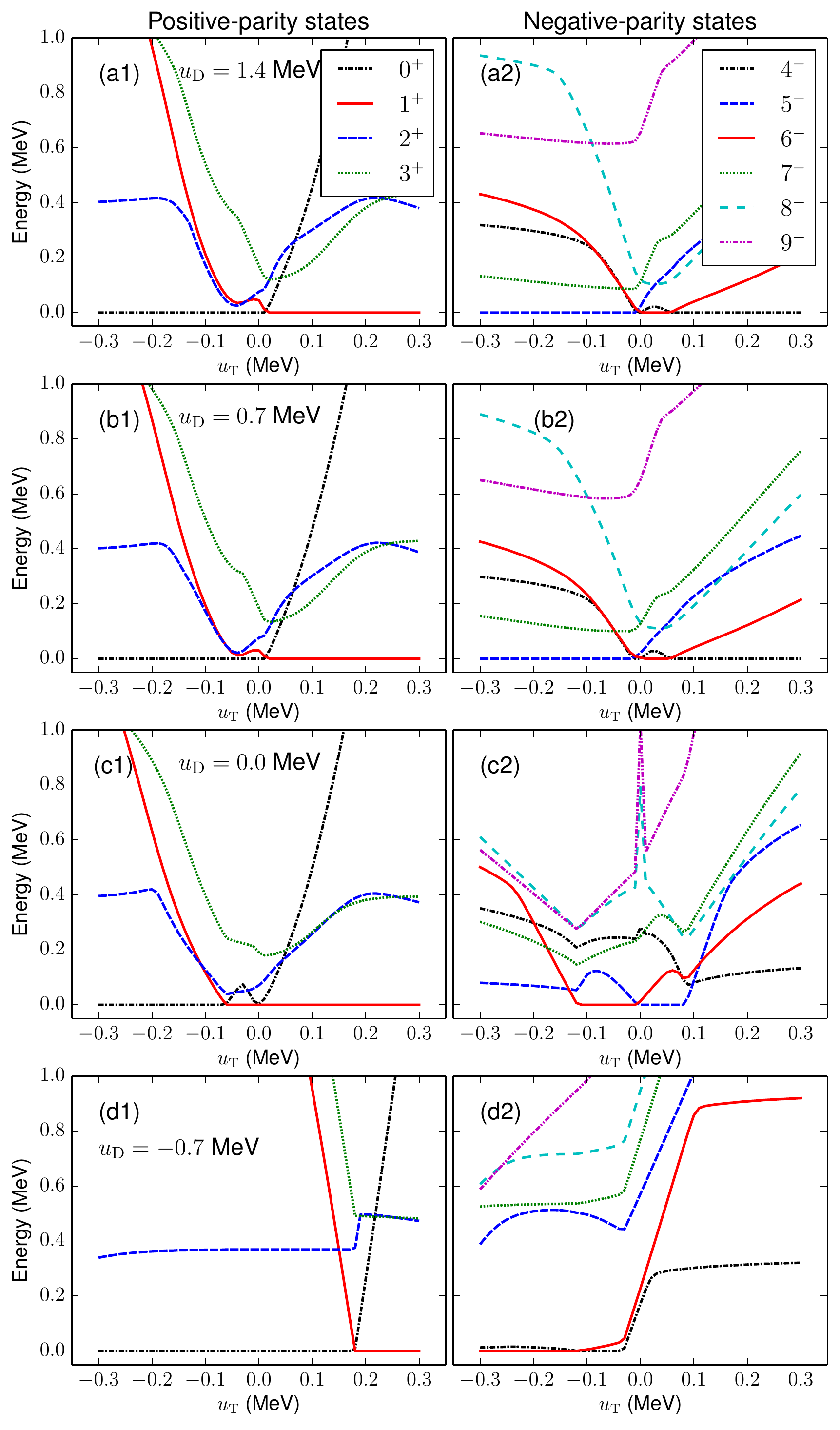}
\caption{(Color online) Excitation energies of the low-lying positive- 
and negative-parity yrast states of the $^{128}$Cs nucleus as functions 
of the parameters $u_\mathrm{T}$ in the cases of different values of the 
parameter $u_\mathrm{D}$, i.e., $u_\mathrm{D}=1.4$ MeV (panels (a1,a2)), 
0.7 MeV (panels (b1,b2)), 0.0 MeV (panels (c1,c2)), and 
$-0.7$ MeV (panels (d1,d2)). 
}
\label{fig:utvd} 
\end{center}
\end{figure*}

\begin{figure}[htb!]
\begin{center}
\includegraphics[width=0.7\linewidth]{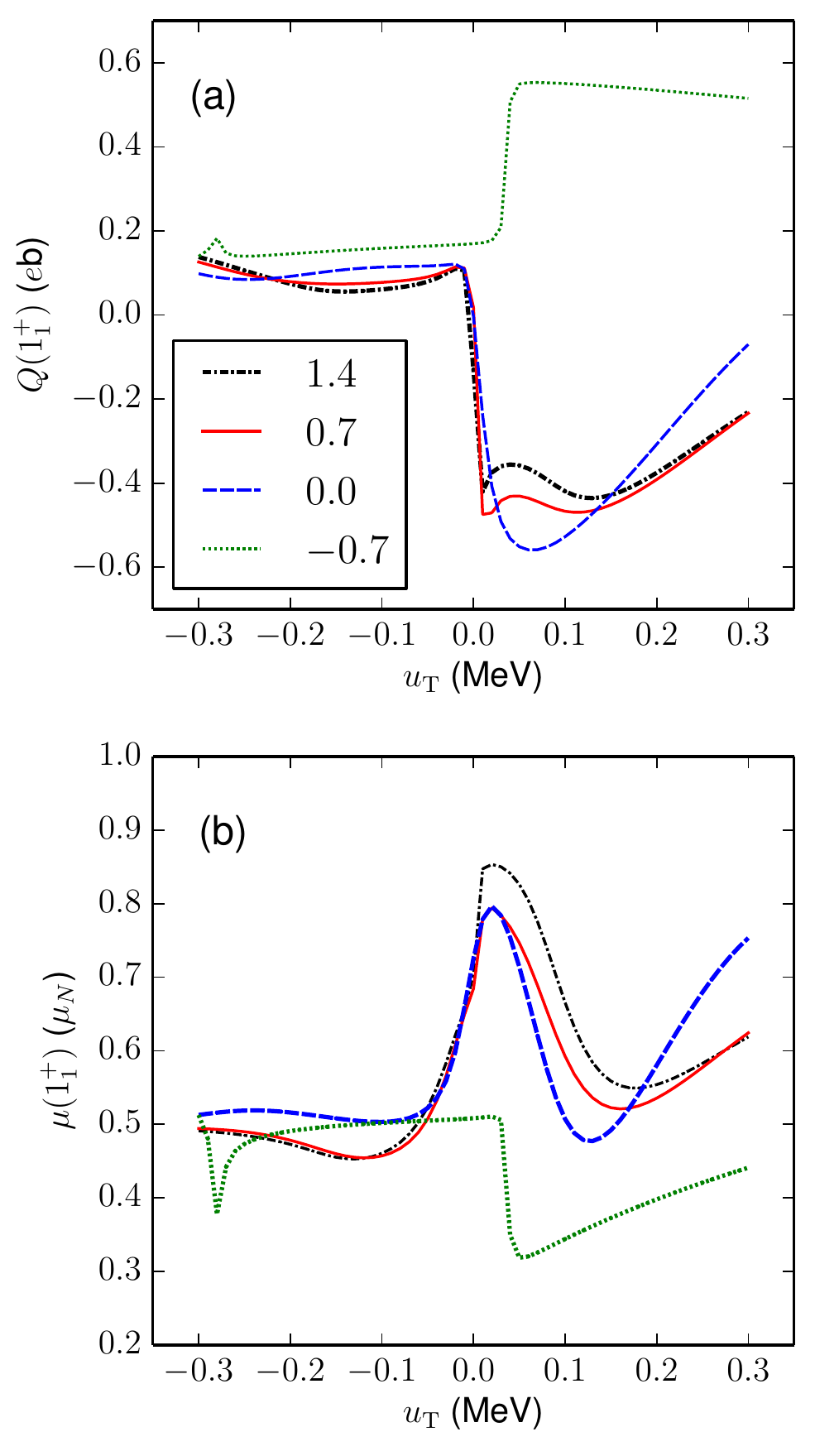}
\caption{(Color online) The calculated quadrupole (a) and magnetic (b) 
moments of the $1^+_1$ ground state for the $^{128}$Cs nucleus as functions 
of the parameters $u_\mathrm{T}$ in the cases of different values of the 
parameter $u_\mathrm{D}$, i.e., $u_\mathrm{D}=1.4$, 0.7, 0.0, and 
$-0.7$ MeV. }
\label{fig:ut-mom} 
\end{center}
\end{figure}


\subsubsection{Band structure of higher-spin states}


We have further studied the detailed band structure of the higher-lying 
higher-spin states in the considered odd-odd Cs isotopes. We have paid 
special attention to the possible doublet structure expected as a 
result of the coupling between a neutron hole and a proton in the 
unique-parity $1h_{11/2}$ orbital. 
Our calculations suggest that the higher-spin states in 
most of the considered odd-odd Cs nuclei are almost entirely composed of 
$[(\nu h_{11/2})^{-1}\otimes\pi h_{11/2}]^{(J)}$ neutron-proton pairs 
coupled to the even-even boson core, as  expected empirically. 

The high-spin bands predicted for the nuclei 
$^{124,126,128,130,132}$Cs are depicted from 
Figs.~\ref{fig:124cs-hs} to \ref{fig:132cs-hs}, respectively. 
In each of these figures, for both theoretical and 
experimental states, the two bands on the left-hand side 
with the $\Delta I=1$ level sequence, and the other two bands 
on the right-hand side with the $\Delta I=1$ level sequence are 
identified as the yrast and side bands, respectively. 
As for the theoretical bands for each nucleus, 
we have simply grouped the calculated 
states $I_1$ (the lowest states with spin $I$) and $I_2$ (the second lowest states 
with spin $I$) into the yrast and side bands, respectively. 

In the case of $^{124}$Cs (see, 
Fig.~\ref{fig:124cs-hs}) the experimental band structure is well 
reproduced, including the energies of the band-head states. 
However, for $^{126}$Cs (see, Fig.~\ref{fig:126cs-hs}) the 
band-head energies of the experimental bands are overestimated 
by a factor of around two. 
Also, because of the limited size of the boson space as the $N=82$  shell closure   
is approached the theoretical bands look more stretched than the 
experimentally identified ones as the spin increases. 
Nevertheless, for 
$^{126}$Cs the overall structure of the theoretical spectrum agrees 
reasonably well with the experimental one. 
The calculated higher-spin bands for $^{126}$Cs shown in Fig.~\ref{fig:126cs-hs} 
resemble  well  the doublet-like bands, i.e., close-lying states with the 
same spin $I$. However, the main 
components of the wave functions of the yrast states up to $I\leq 17^+$ 
are coming from the coupling between the odd neutron and odd proton 
in the normal-parity $sdg$ orbitals, not the 
$[(\nu h_{11/2})^{-1}\otimes\pi h_{11/2}]^{(J)}$ 
neutron-proton pair configurations as in all the other odd-odd Cs 
nuclei considered. 

As for $^{128}$Cs, the absolute energies of the observed 
bands have not been established experimentally. 
Therefore, in Fig.~\ref{fig:128cs-hs} 
both the experimental and calculated energy levels for $^{128}$Cs are 
plotted with respect to the experimental $10^+_1$ state, which is 
suggested to be the band-head of the lowest-energy band based on the 
$[(\nu h_{11/2})^{-1}\otimes\pi h_{11/2}]^{(J)}$ configuration. In 
general, the structures of the bands identified experimentally are well 
reproduced up to $I\approx 16^+$. However, the energy of higher-spin 
states is overestimated. With increasing spin the stretching of the 
predicted bands, as compared with the experiment, becomes larger than 
in $^{126}$Cs.
 
In Figs.~\ref{fig:130cs-hs} and \ref{fig:132cs-hs}, similar doublet-like 
band structures are obtained also in $^{130,132}$Cs and are in a good 
agreement with the experimental spectra up to relatively low spin, 
e.g., $I\leq 15^+$. The moments of inertia for higher spin states 
are considerably underestimated due to the fact that the configuration 
space of the even-even boson core becomes much smaller for those 
nuclei close to the neutron shell closure $N=82$. 
For instance, in $^{132}$Cs there is only one $19^+$ state, 
which is formed by the configuration where four $d$ bosons and 
a neutron and a proton in the $h_{11/2}$ orbital are all aligned, i.e., 
$L=8$ and $[(\nu h_{11/2})^{-1}\otimes\pi h_{11/2}]^{(11)}$. 
Also there is no state with spin higher than $I=19^+$. 
Possible solutions to improve the description of the higher-spin states 
in the IBFFM-2 can be, for instance, the inclusions of an additional 
boson degree of freedom, e.g., $L=4^+$ ($g$) boson, 
and of higher quasiparticle excitations or broken pairs. 
These extensions are, however, out of the scope of the present study. 
 

\subsubsection{$B(E2)$ and $B(M1)$ systematic in the high-spin states}


To identify possible signatures of chirality we have considered, in 
addition to energy levels, the systematic of the E2 and M1 transitions 
with increasing spin. Our analysis of the $B(E2)$ and $B(M1)$ patterns 
suggests that there are many examples in the odd-odd Cs nuclei 
that can be considered candidates to display chirality. 
In particular, the observed $B(E2; I\rightarrow I-2)$ and $B(M1; 
I\rightarrow I-1)$ intra-band and inter-band transitions in the yrast 
and second-lowest bands of the $^{128}$Cs nucleus show a definite staggering 
pattern as a function of the angular momentum \cite{grodner2006}. Such 
a selection rule has been derived from  symmetry considerations applied 
to a simple particle-rotor model \cite{koike2004}. Nevertheless, they 
can still  be used to benchmark our calculations. 

The predicted $B(E2; I\rightarrow I-2)$ transition rates for most of 
the considered double-odd nuclei $^{124-132}$Cs (see, 
panels (a1) to (e1) on the left-hand side of 
Fig.~\ref{fig:em-hs}) do not show any such staggering as the one 
that appears in the simplified model \cite{koike2004}. For the yrast 
band they evolve monotonously or stay rather constant with $I$. 
In some of the $B(E2)$ transitions shown, at particular spin their 
values almost vanish, e.g., the $B(E2; 16^+_2\rightarrow 14^+_2)$ 
transition rate in $^{128}$Cs in panel (c1). 
Particularly irregular $I$-dependence of the predicted $B(E2)$ rates 
is found for the $^{126}$Cs nucleus (see, panel (b1)). 
This is because certain mixing among states 
with a given spin tends to occur and, therefore, the assignments of the 
lowest states $I_1$ into the yrast band and of 
the second-lowest states $I_2$ into 
the side band are, in some cases, not adequate. 
In a number of the odd-odd Cs nuclei, however, a certain staggering pattern, 
similar to the one in the observed $B(M1; I\rightarrow I-1)$ rates for $^{128}$Cs  
\cite{grodner2006}, has been obtained in the calculated 
$B(M1; I\rightarrow I-1)$ rates for both the intra- (middle panels (a2) to (e2) of 
Fig.~\ref{fig:em-hs}) and inter-band (right panels (a3) to (e3)) transitions.


As yet another indication of the chiral bands, we show in 
Fig.~\ref{fig:em-ratio} the ratio of the calculated $B(M1; I\rightarrow I-1)$ 
to $B(E2; I\rightarrow I-2)$ rates of the yrast and side bands for 
all the odd-odd Cs nuclei. 
A number of experimental values for these quantities
are available in Refs.~\cite{gizon2001,wang2006,paul1989,simons2005,rainovski2003,xiong2019}. 
Our results show
a staggering pattern of the $B(M1)/B(E2)$ ratio as a function of 
angular momentum $I$ for both yrast and side bands 
in all the considered odd-odd Cs nuclei, except $^{124}$Cs (panel (a1) of Fig.~\ref{fig:em-ratio}).  
The theoretical values are consistent with 
the empirical trend (shown on the right-hand side of Fig.~\ref{fig:em-ratio}). 
However, the predicted $B(M1)/B(E2)$ values are much larger in magnitude
and show a more irregular $I$-dependence than the experimental data.
The quantitative disagreement can be expected from the behavior of the calculated 
$B(E2; I_{1,2}\rightarrow (I-2)_{1,2})$ (Figs.~\ref{fig:em-hs}(a1-e1))
and $B(M1; I_{1,2}\rightarrow (I-1)_{1,2})$ (Figs.~\ref{fig:em-hs}(a2-e2)) 
values as functions of $I$. 

In order to examine whether the predicted yrast and side bands can be considered 
partners of the chiral doublet, we show in Fig.~\ref{fig:mom} the 
quadrupole moment $Q(I)$ (in $e$b units) and the $g$-factor for 
the corresponding states in the considered odd-odd $^{124-132}$Cs nuclei. 
The $Q(I)$ values are negative and decrease in magnitude with increasing spin. 
In addition, we have obtained similar $Q(I)$ values and $I$-dependence, 
for both bands. 
The above observation does not apply to the results for the 
$^{126}$Cs nucleus in panel (b1). 
This is because, as we mentioned earlier, 
the states in the yrast and side bands for $^{126}$Cs in particular, predicted 
in the present calculation, have different wave function contents. 
Namely, the side-band states are mainly composed of the 
$[(\nu h_{11/2})^{-1}\otimes\pi h_{11/2}]^{(J)}$ single-particle 
configuration, but this is not the case for those states 
in the yrast band. 
In all the odd-odd Cs nuclei, the $g$-factor values, 
depicted on the right-hand side of the same figure, 
are quite similar (around 0.5) for both bands.
Note, that the $g$-factor obtained for the 
$I=9^+$ yrast state for $^{128}$Cs agrees well with the experimental value 
($+0.59\pm 0.01$) \cite{grodner2018}.


\subsubsection{Dependence on the residual neutron-proton interaction}


Finally we examine how the spectroscopic results from the IBFFM-2 
depend on the choice of the strength parameters for the residual 
neutron-proton interaction $\hat V_\mathrm{res}$ (see, Eq.~(\ref{eq:res})). 
In Fig.~\ref{fig:utvd} we depict the evolution of the calculated excitation energies 
for a few low-lying positive- (panels (a1)--(d1) on the left-hand side 
of Fig.~\ref{fig:utvd}) and negative-parity 
(panels (a2)--(d2) on the right-hand side) 
yrast states of the $^{128}$Cs nucleus as functions of the strength parameter 
for the tensor interaction $u_\mathrm{T}$ (in MeV), in the cases of different values of the 
parameter for the delta interaction, i.e., $u_\mathrm{D}=1.4$ MeV 
(panels (a1,a2)), 0.7 MeV (panels (b1,b2)), 
0.0 MeV (panels (c1,c2)), and $-0.7$ MeV (panels (d1,d2)). 
The calculated excitation energies, especially for the positive-parity ones, 
do not seem to have a strong dependence on the parameter $u_\mathrm{D}$ 
for $u_\mathrm{D}>0$ MeV, but are rather sensitive to $u_\mathrm{T}>0$ 
MeV. 
Our choice, i.e., the value of the parameter $u_\mathrm{T}$, that is somewhere in the range 
$0< u_\mathrm{T}<0.05$, as well as the value of $u_\mathrm{D}$ to be approximately 
0.7 MeV, seems to be optimal for reproducing the ground-state spin for both the 
positive- and negative-parities, i.e., $I=1^+_1$ and $6^-_1$, respectively. 

Similarly, in Fig.~\ref{fig:ut-mom} we plot the 
calculated quadrupole $Q(1^+_1)$ and magnetic $\mu(1^+_1)$ 
moments for the ground state $1^+_1$ of the $^{128}$Cs nucleus 
as functions of the strength parameter $u_\mathrm{T}$, in the 
cases of different values of the 
parameter $u_\mathrm{D}=1.4$, 0.7, 0.0, and $-0.7$ MeV. 
Both the $Q(1^+_1)$ and $\mu(1^+_1)$ values depend somewhat largely 
on the parameter $u_\mathrm{T}$ with $u_\mathrm{T}>0$ MeV, 
but are much less sensitive to the parameter $u_\mathrm{D}>0$ MeV. 
The chosen value for the tensor interaction strength $u_\mathrm{T}$ of 
0.02 MeV gives both the $Q(1^+_1)$ and $\mu(1^+_1)$ values 
close to the corresponding experimental data, $Q(1^+_1)=-0.570\pm 0.08$ $e$b 
and $\mu(1^+_1)=+0.974\pm 0.005$ $\mu_N$, respectively. 

Similar parameter dependences of the excitation energies and moments 
have been obtained for the other odd-odd Cs studied here. 
In principle, one could use the values of the $u_\mathrm{D}$ and $u_\mathrm{T}$ 
parameters different from nucleus to nucleus and/or between both parities. 
We have used the fixed values for the $u_\mathrm{D}$ and $u_\mathrm{T}$ 
parameters only for the sake of simplicity. 
Nevertheless, we consider the chosen values of the 
$u_\mathrm{D}$ and $u_\mathrm{T}$ strength parameters realistic 
in a sense that the overall description 
of the energy levels of the low-lying low-spin states for the studied 
odd-odd nuclei is reasonable.


\section{Summary and concluding remarks\label{sec:summary}}


The spectroscopic properties of the odd-odd nuclei $^{124-132}$Cs have 
been analyzed using the interacting boson-fermion-fermion (IBFFM-2) 
framework with microscopic input from mean field calculations with the 
Gogny-D1M energy density functional. The $(\beta,\gamma)$-deformation 
energy surface  for  even-even boson-core Xe isotopes as well as 
single-particle energies and occupation probabilities of unpaired 
nucleons in odd-N Xe, odd-Z Cs and odd-odd Cs nuclei obtained from the 
mean field calculation are used to build, via a mapping procedure, the 
corresponding IBFFM-2 Hamiltonian. In its current implementation, the 
method still requires a few coupling constants of the boson-fermion and 
residual neutron-proton interactions to be fitted to the experiment. 
The diagonalization of the corresponding IBFFM-2 Hamiltonian provides 
wave functions, energy levels as well as other spectroscopic properties 
such as E2 and M1 transition rates.

It has been shown, that the (mapped) IBFFM-2 model describes reasonably 
well both the positive- and negative-parity low-lying low-spin states 
of the considered odd-odd Cs nuclei, especially in the case of 
$^{124,126,128}$Cs. This is a remarkable result, considering the 
significant reduction of parameters with respect to previous IBFFM 
calculations. However, in some of the odd-odd nuclei (e.g., in 
$^{132}$Cs) the ordering of both the positive- and negative-parity 
levels close to the ground state could not be correctly reproduced by 
our calculations. Some possible explanations for this failure are the 
limited number of active bosons in the even-even core near the shell 
closure; the possibility that the adopted single-particle energies and 
occupation numbers provided by the Gogny-D1M HFB approach may not be 
realistic enough; finally the use of fixed strengths for the whole 
isotopic chain for the residual neutron-proton interaction in the 
IBFFM-2 Hamiltonian. 

We have also studied the band structure of the higher-spin 
positive-parity states in the considered odd-odd Cs nuclei. 
Our calculations provide 
a reasonable quantitative description of the excitation energies of 
these bands up to $I\approx 20^+$ except for  the excitation energy of 
the band-heads of  $^{126}$Cs which are overestimated. 
We have identified many of the double-odd Cs nuclei as good 
candidate for the existence of chiral doublet bands. 
In particular, the calculated $B(M1; I\rightarrow I-1)$ transition rates 
exhibit staggering patterns with increasing angular momentum. 
This result agrees well  with the selection rule derived by  simple symmetry 
considerations \cite{koike2004}. All in all, the results of this study 
suggest that the employed theoretical methods can be potentially used 
to describe even such a type of nuclear excitation as chirality.

\acknowledgments
The work of KN is financed within the Tenure Track Pilot Program of the Croatian Science 
Foundation and the \'Ecole Polytechnique F\'ed\'erale de Lausanne and the Project TTP-2018-07-3554
Exotic Nuclear Structure and Dynamics, with funds of the Croatian-Swiss Research Program.
the Croatian Science Foundation and 
\'Ecole Polytechnique F\'ed\'erale de Lausanne under the Swiss-Croatian 
Corporation Program No. TTP-2018-07-3554. 
The  work of LMR was supported by the Spanish Ministry of Economy and Competitiveness (MINECO)
Grants No. FPA2015-65929-MINECO and FIS2015-63770-MINECO.

\bibliography{refs}

\end{document}